\documentclass[journal]{IEEEtran}
\usepackage{amsmath,amsfonts}
\usepackage{algorithmic}
\usepackage{algorithm}
\usepackage{array}
\usepackage[caption=false,font=footnotesize,labelfont=rm,textfont=rm]{subfig}
\usepackage{textcomp}
\usepackage{stfloats}
\usepackage{url}
\usepackage{verbatim}
\usepackage{graphicx}
\usepackage{cite}
\usepackage{color}
\usepackage{amssymb}
\usepackage{amsthm}
\usepackage{cuted}
\usepackage{multirow}  % For multirow cells in the table
\usepackage{bm}
\usepackage{float}
\usepackage{pifont}
\usepackage{enumitem,kantlipsum}
\usepackage{xpatch}     % 用于修补命令
\usepackage{xcolor}     % 用于颜色控制
\usepackage{expl3}      % 用于 expl3 语法
\usepackage{etoolbox}   % 可选，增强宏处理能力
\makeatletter
\ExplSyntaxOn
\cs_new:Npn \bibColoredItems #1#2
  {
    \clist_map_inline:nn {#2} { \cs_new:cpn {bib@colored@##1} {#1} } 
  }
\ExplSyntaxOff

\newcommand\bib@setcolor[1]{%
  \ifcsname bib@colored@#1\endcsname
    \expandafter\color\expandafter{\csname bib@colored@#1\endcsname}
  \else
    \normalcolor
  \fi
}

\xpatchcmd\@bibitem
  {\item}
  {\bib@setcolor{#1}\item}
  {}{\PatchFailed}

\xpatchcmd\@lbibitem
  {\item}
  {\bib@setcolor{#2}\item}
  {}{\PatchFailed}

\makeatother

\newtheorem{Lemma}{Lemma}
\newtheorem{remark}{Remark}

\newcommand{\cmark}{\ding{51}}%
\newcommand{\xmark}{\ding{55}}%

\newcommand{\bfI}{\mathbf{I}}
\newcommand{\bfH}{\mathbf{H}}
\newcommand{\bfV}{\mathbf{V}}
\newcommand{\bfF}{\mathbf{F}}

\newcommand{\bfGamma}{\mathbf{\Gamma}}
\newcommand{\bfLambda}{\mathbf{\Lambda}}

\newcommand{\bfY}{\mathbf{Y}}
\newcommand{\bfL}{\mathbf{L}}
\newcommand{\bfK}{\mathbf{K}}
\newcommand{\bfZ}{\mathbf{Z}}

\newcommand{\bfD}{\mathbf{D}}

\newcommand{\bfX}{\mathbf{X}}

\begin{document}

\title{DeepFP: Deep-Unfolded Fractional Programming for MIMO Beamforming}

\author{Jianhang Zhu,~\IEEEmembership{Graduate Student Member,~IEEE}, Tsung-Hui Chang,~\IEEEmembership{Fellow,~IEEE},\\
Liyao Xiang,~\IEEEmembership{Member,~IEEE}, and Kaiming Shen,~\IEEEmembership{Senior Member,~IEEE}
% \thanks{This paper was produced by the IEEE Publication Technology Group. They are in Piscataway, NJ.}% <-this % stops a space
% \thanks{Manuscript received April 19, 2021; revised August 16, 2021.}
\thanks{Manuscript accepted to IEEE Transactions on Communications on January 3, 2026. The work of Jianhang Zhu and Kaiming Shen was supported in part by the NSFC under Grant 12426306 and in part by Guangdong Basic and Applied Basic Research under Grant 2023B0303000001. The work of Liyao Xiang was supported by the NSFC under Grant U25A20445 and Grant 62272306. An earlier version of this paper has been presented in part at IEEE SPAWC 2025 \cite{Zhu2025DeepFP}. Source codes are available at https://github.com/zhujhz/DeepFP.git. \emph{(Corresponding author: Kaiming Shen.)}}
\thanks{
    Jianhang Zhu and Kaiming Shen are with the School of Science and Engineering, The Chinese University of Hong Kong (Shenzhen), 518172 Shenzhen, China (e-mail: jianhangzhu1@link.cuhk.edu.cn; shenkaiming@cuhk.edu.cn).}
\thanks{Tsung-Hui Chang is with the School of Artificial Intelligence, The Chinese University of Hong Kong (Shenzhen), 518172 Shenzhen, China (e-mail: tsunghui.chang@ieee.org).}
\thanks{Liyao Xiang is with Shanghai Jiao Tong University, 200240, Shanghai, China (e-mail: xiangliyao08@sjtu.edu.cn).}

}
\maketitle

\begin{abstract}
This work proposes a mixed learning-based and optimization-based approach to the weighted-sum-rates beamforming problem in a multiple-input multiple-output (MIMO) wireless network. The conventional methods, i.e., the fractional programming (FP) method and the weighted minimum mean square error (WMMSE) algorithm, can be computationally demanding for two reasons: (i) they require inverting a sequence of matrices whose sizes are proportional to the number of antennas; (ii) they require tuning a set of Lagrange multipliers to account for the power constraints. The recently proposed method called the reduced WMMSE addresses the above two issues for a single cell. In contrast, for the multicell case, another recent method called the FastFP eliminates the large matrix inversion and the Lagrange multipliers by using an improved FP technique, but the update stepsize in the FastFP can be difficult to decide. As such, we propose integrating the deep unfolding network into the FastFP for the stepsize optimization. Numerical experiments show that the proposed method is much more efficient than the learning method based on the WMMSE algorithm.
\end{abstract}


\section{Introduction}

A fundamental problem of multiple-input-multiple-output (MIMO) system design is to optimize the transmit beamformers to maximize the weighted-sum-rates (WSR) throughout cellular networks, namely the WSR problem. The weighted minimum mean square error (WMMSE) algorithm \cite{Shi2011WMMSE,Christensen2008WMMSE} and the fractional programming (FP) \cite{kaiming2018FPI,kaiming2018FP2} constitute two popular approaches in this area. Since two methods are both iteratively structured, a natural idea is to learn their behaviors via deep unfolding, as pursued extensively in \cite{Hu2021IAIDNN, Minghe2023Learning, Pellaco2022UFWMMSE,Nguyen2023DUHybridbeamforming,Hu2021DURLMIMO,Xu2023JointSchebeamforming,Schynol2023DUMIMO,Chowdhury2024GNNMIMO}. However, two main challenges arise when it comes to the multicell MIMO case: (i) the iterative algorithm (e.g., WMMSE or FP) requires inverting large matrices, yet the matrix inversion is much more difficult to learn than the matrix addition and multiplication; (ii) the iterative algorithm requires finding the optimal Lagrange multipliers for the power constraint, which is complicated and highly nonlinear and can increase the training cost considerably. To address these issues, this paper proposes a novel deep unfolding scheme called \emph{DeepFP} that relies on the new FP technology. Differing from the previous methods \cite{Hu2021IAIDNN, Minghe2023Learning, Pellaco2022UFWMMSE,Nguyen2023DUHybridbeamforming,Hu2021DURLMIMO,Xu2023JointSchebeamforming,Schynol2023DUMIMO,Chowdhury2024GNNMIMO}, DeepFP avoids learning the large matrix inversion and the nonlinear optimization of Lagrange multipliers, and only focuses on how to coordinate a small set of scalar stepsizes. Here is a big picture of how this work is developed. The original objective function for the beamforming problem is $f_o$, which is difficult to tackle directly. The conventional FP method \cite{kaiming2018FPI,kaiming2018FP2} suggests converting $f_o$ to $f_q$ so that the iterative optimization is easy to perform, but then the new issue is that it entails computing the large matrix inverse. To get rid of this complexity, a more recent method called the nonhomogeneous quadratic transform \cite{Kaiming2024NFP} further converts $f_q$ to $f_n$, but then its performance is sensitive to the choice of stepsize: if the stepsize is too large, then the iteration may not converge; if the stepsize is too small, then the convergence would slow down. To decide the stepsize, \cite{Kaiming2024NFP} imposes a strong assumption that all the users in the same cell use the same stepsize, and then shows that the stepsize is upper bounded by some eigenvalue. In contrast, this paper considers tuning the stepsize separately for each individual user across the network, so this eigenvalue-based upper bound disappears, and thus the choice of stepsize can be more aggressive.

The WSR problem is notoriously difficult. In fact, it is shown to be NP-hard even for the single-input-single-output case \cite{luo2008Dynamic}. Aside from the branch-and-bound approaches in \cite{Joshi2011WSRBB, Liu2012WSRBB}, most existing works aim to find a local optimum efficiently. The classic methods include the maximum ratio transmission (MRT) \cite{Kammoun2014MRT}, the zero-forcing (ZF) method \cite{Gao2011ZF}, and the regularized ZF precoding (RZF) method \cite{Nguyen2019RZF}, which are verified at the link level under certain conditions but can lead to quite large performance losses at the system level. In more recent literature, the WMMSE algorithm \cite{Shi2011WMMSE,Christensen2008WMMSE} is widely adopted for solving the WSR problem. Its main idea is to utilize a connection between the rate maximization and the mean square error (MSE) minimization to rewrite the WSR problem as a weighted MSE minimization problem---which can be efficiently solved by the block coordinate descent (BCD) method~\cite{Bertsekas1999} in an iterative fashion. Thanks to the BCD theory, the WMMSE algorithm has provable convergence to a stationary point solution of the WSR problem.

However, the WMMSE algorithm can incur high computational tension in the multicell MIMO case. To be more specific, each iterate of WMMSE requires inverting a matrix whose size is proportional to the number of transmit antennas. Thus, in the multicell MIMO case with a large number of antennas deployed at the transmitter side, the WMMSE algorithm requires lots of large matrix inversions. Such tension has been relieved more or less by a recent work \cite{Xiaotong2023RWMMSE}. The main idea of \cite{Xiaotong2023RWMMSE} is to recast the beamforming vectors to a new space whose dimension only depends on the total number of receive antennas (or the number of users, assuming each user has only one receive antenna). This modified WMMSE algorithm (referred to as the RWMMSE in \cite{Xiaotong2023RWMMSE}) now instead inverts matrices whose sizes are proportional to the number of users. Clearly, the RWMMSE algorithm has reduced complexity only when there are a limited number of users in the network.

\begin{table*}[t]
\renewcommand\arraystretch{1.2}
        \centering
        \caption{Comparison of the different methods for MIMO beamforming when each transmitter has $N$ antennas.}\label{tab:algorithm compare}
        \begin{tabular}{|c|c|c|c|c|c|c|}
            \hline
                Method & Can avoid $N\times N$ matrix inversion?& Can avoid Lagrange multipliers tuning? & Can work for multiple cells? \\
            \hline
            \hline
            WMMSE \cite{Shi2011WMMSE,Christensen2008WMMSE}  & \xmark & \xmark & \cmark \\
            \hline
            Reduced WMMSE \cite{Xiaotong2023RWMMSE} & \xmark\;(but can reduce the matrix size) & \cmark & \xmark \\
            \hline
            FP \cite{kaiming2018FPI} & \xmark & \xmark & \cmark \\
            \hline
            Deep unfolding + FP \cite{Hu2021IAIDNN} & \xmark & \cmark & \cmark \\
            \hline
            FastFP \cite{Kaiming2024NFP,Zhang2023Enhangcing} & \cmark\;(but requires eigencomputation) & \cmark & \cmark \\
            \hline
            Proposed DeepFP & \cmark & \cmark & \cmark\\
            \hline
        \end{tabular}
\end{table*}

Another challenge faced by the WMMSE algorithm in multicell MIMO is caused by the power constraint. Specifically, in each iteration, WMMSE needs to determine a Lagrange multiplier for each cell to satisfy the power constraint on the beamforming vectors. The optimal Lagrange multiplier has no closed-form solution and is typically addressed via bisection search~\cite{Shi2011WMMSE}. The recently proposed ``reduced WMMSE'' algorithm \cite{Xiaotong2023RWMMSE} has partially addressed this issue. The authors of \cite{Xiaotong2023RWMMSE} show that it is optimal to scale all the beamforming vectors simultaneously to meet the power constraint when considering a single cell. However, it is difficult to extend the above result for multiple cells. Another approach to the multicell MIMO beamforming problem is based on the manifold optimization \cite{Rui2024Manifold}. Its main idea is to restrict the beamforming variables to a Riemannian manifold defined by the power constraint, thereby converting the constrained optimization to the unconstrained. However, the manifold method only optimizes beamforming vectors under the fixed power levels, whereas WMMSE can optimize beamforming vectors and powers jointly. Moreover, its performance is verified only for the single-cell network.

Aside from the above model-driven method, there is a surge of research interest in the data-driven approach to the multicell MIMO beamforming problem. Differing from those pure black-box learning methods \cite{Haoran2018L2O,wei2018DL,wenchao2020DLMISO,hengtao2018DLMIMO} that attempt to mimic the existing optimization methods (e.g., WMMSE) via the universal approximation of capability deep neural network (DNN), the deep unfolding methods \cite{Gregor2010LISTA,hershey2014deepunfolding,Stimming2019DUCom} take into account the iterative structure of the conventional model-driven algorithms and aim to learn the behavior of each iteratation. For the WSR beamforming problem, the previous studies \cite{Hu2021IAIDNN, Minghe2023Learning, Pellaco2022UFWMMSE,Nguyen2023DUHybridbeamforming,Hu2021DURLMIMO,Xu2023JointSchebeamforming,Schynol2023DUMIMO,Chowdhury2024GNNMIMO} mostly take the WMMSE algorithm as the learning target of deep unfolding. For example, the deep unfolding network in \cite{Hu2021IAIDNN} aims at the RWMMSE algorithm, while \cite{Pellaco2022UFWMMSE,Minghe2023Learning} aim at the WMMSE algorithm. However, as these deep unfolding methods successfully mimic WMMSE, they in the meanwhile inherit the aforementioned drawbacks of their target algorithms. As such, the deep unfolding network in \cite{Hu2021IAIDNN} can only handle a single cell, \cite{Pellaco2022UFWMMSE} has to approximate the large matrix inversion in a suboptimal approximate fashion, and \cite{Minghe2023Learning} is limited to the multiple-input-single-output (MISO) case in order to avoid learning the bisection search for the optimal Lagrange multipliers.

To overcome the above bottleneck, the deep unfolding method proposed in this paper takes advantage of an intimate connection between WMMSE and FP. Roughly speaking, FP refers to a class of optimization problems which are fractionally structured, e.g., the sum-of-ratios maximization. It turns out that the WSR problem can be recast to a sum-of-ratios problem, and accordingly the WMMSE algorithm boils down to a special case of the FP algorithm\cite{kaiming2018FPI,kaiming2018FP2}. In fact, the large matrix inversion and the Lagrange multiplier optimization have been well studied in the realm of FP, e.g., the so-called \emph{nonhomogeneous quadratic transform} \cite{Kaiming2024NFP,Zhang2023Enhangcing} can address both issues. Thus, unlike the previous works \cite{Hu2021IAIDNN, Minghe2023Learning, Pellaco2022UFWMMSE,Nguyen2023DUHybridbeamforming,Hu2021DURLMIMO,Xu2023JointSchebeamforming,Schynol2023DUMIMO,Chowdhury2024GNNMIMO} that consider deep-unfolding the WMMSE algorithm directly, this work proposes incorporating the inhomogeneous quadratic transform into the deep unfolding paradigm. We then show that the core of the learning task is to decide the stepsize used in the inhomogeneous quadratic transform-based FP. The main features and advantages of the proposed method are summarized in Table \ref{tab:algorithm compare}. Our work introduces two main novelties:
\begin{itemize}
\item \emph{A New Paradigm of Deep-Unfolding:} The existing deep-unfolding methods for beamforming typically mimic the behavior of the traditional FP (based on the quadratic transform) or the WMMSE algorithms. In contrast, this work builds upon the nonhomogeneous quadratic transform, which inherently avoids the complexity of computing the large matrix inverse. Furthermore, unlike the existing nonhomogeneous quadratic transform in \cite{Kaiming2024NFP} that tunes the stepsize parameter on a per-cell basis, this work novelly suggests optimizing the stepsize parameter for each individual user. The numerical results show that this new paradigm leads to faster convergence and superior performance.
\item \emph{A New Hybrid Training Strategy:} The proposed DeepFP employs a novel hybrid training strategy that combines supervised and unsupervised learning. Specifically, in the first stage, the DNN is initialized and trained using model-driven FastFP solutions as labels to ensure rapid convergence and robust initialization. Subsequently, in the second stage, the DNN is fine-tuned using the actual weighted sum-rate objective as the loss function, allowing the network to refine its performance toward the true objective.
\end{itemize}

The remainder of this paper is organized as follows. Section \ref{sec:problem} introduces the weight sum-rate problem formulation. Section \ref{model-Driven Methods} shows and compares existing model-driven algorithms, including the FP algorithm and the FastFP algorithm. Section \ref{sec:unfolding} develops our proposed DeepFP network based on the FastFP algorithm. Section \ref{sec:results} presents numerical results. Finally, Section \ref{sec:conclusion} concludes the paper. 

% , and $\|\mathbf{a}\|_2$ is its $\ell_2$ norm
% $\mathbf{A}^*$ is its complex conjugate, $\mathbf{A}^\top$ is its transpose,
% $[\mathbf{A}]_{mm}$ is the $m$th  diagonal element of $\mathbf{A}$.

Here and throughout, bold lower-case letters represent vectors while bold upper-case letters represent matrices. For a vector $\mathbf{a}$, $\mathbf{a}^H$ is its conjugate transpose. For a matrix $\mathbf{A}$, $\mathbf{A}^H$ is its conjugate transpose and $\|\mathbf{A}\|_F$ is its Frobenius norm.  $\text{col}(\mathbf{A})$ refers to the number of columns in matrix $A$. For a square matrix $\mathbf{A}$, $\text{tr}(\mathbf{A})$ is its trace, $|\mathbf{A}|$ is its determinant, and $\lambda_{\max}(\mathbf{A})$ is its largest eigenvalue. Denote by $\mathbf{I}_d$ the $d \times d$ identity matrix, $\mathbb{C}^n$ the set of $n \times 1$ vectors, $\mathbb{C}^{d \times n}$ the set of $d \times n$ matrices, and $\mathbb{H}_+^{d \times d}$ the set of $d \times d$ positive definite matrices. For a complex number $\mathbf{a} \in \mathbb{C}$, $\Re\{\mathbf{a}\}$ is its real part, $|\mathbf{a}|$ is its absolute value. The underlined letters represent the collections of the associated vectors or matrices, e.g., for $\mathbf{a}_1, \dots, \mathbf{a}_n \in \mathbb{C}^d$ we write $\underline{\mathbf{a}}  = [\mathbf{a}_1, \mathbf{a}_2, \dots, \mathbf{a}_n]^\top \in \mathbb{C}^{n\times d}$.

\begin{figure*}[htbp] 
        \centering
        \begin{align}\label{equ:MSE_q}
                f_q(\underline{\bfV},\underline{\bfGamma},\underline{\bfY}) = 
                \sum_{\ell,k} \Big[ \operatorname{tr} \left( 2\Re \left\{ \mathbf{V}_{\ell k}^H \mathbf{\Lambda}_{\ell k} \right\} 
                - \omega_{\ell k} \mathbf{Y}_{\ell k}^H \mathbf{D}_{\ell k} \mathbf{Y}_{\ell k} \left(  \mathbf{I}_{d}+\bfGamma_{\ell k} \right)  \right)
                + \omega_{\ell k} \log \left| \mathbf{I}_{d}+\bfGamma_{\ell k} \right| 
                - \operatorname{tr} \left( \omega_{\ell k} \mathbf{\Gamma}_{\ell k} \right) \Big]\tag{7}
        \end{align}
        \hrulefill
        % \vspace*{2pt}
\end{figure*}

\section{Weighted Sum-Rate Maximization Problem}\label{sec:problem}
Consider a downlink multi-user multiple-input-multiple-output (MU-MIMO) system with $L$ cells. Within each cell, one base station (BS) with $N_{t}$ transmit antennas serves $K$ users. The $k$th user in the $\ell$th cell is indexed as $(\ell,k)$. Assume that user $(\ell,k)$ has $ N_r $ receive antennas and that $ d $ data streams are intended for it. Let $ \mathbf{V}_{\ell k} \in \mathbb{C}^{N_t \times d} $ represent the beamforming matrix used by BS $ \ell $ associated with the signal $ \mathbf{s}_{\ell k} \in \mathbb{C}^{d \times 1} $ for user $ (\ell, k) $. Assuming that $
\mathbb{E}[\mathbf{s}_{\ell k} \mathbf{s}_{\ell k}^{H}] = \mathbf{I}_{d}$, the received signal $y_{\ell k}$ at user $(\ell ,k)$ is given by

\begin{align}
        \mathbf{y}_{\ell k} = \underbrace{\bfH_{\ell k,\ell} \bfV_{\ell k} \mathbf{s}_{\ell k}}_{\text{desired signal}} &+ \underbrace{\sum_{j=1,j \neq k}^{K} \bfH_{\ell k,\ell} \bfV_{\ell j} \mathbf{s}_{\ell j}}_{\text{intracell interference}}\nonumber\\
        &+ \underbrace{\sum_{i=1, i \neq \ell}^{L} \sum_{j=1}^{K} \bfH_{\ell k,i} \bfV_{ij} \mathbf{s}_{ij} }_{\text{intercell interference}} + \mathbf{n}_{\ell k},
\end{align}
where the channel state information (CSI) $\bfH_{\ell k,i} \in \mathbb{C} ^{ N_r \times N_t}$ is the channel from BS $i$ to user $(\ell,k)$, and $\mathbf{n}_{\ell k}\sim\mathcal{CN} (\bm0,\sigma^2\mathbf{I})$ is the additive white Gaussian noise with power level $\sigma^2$. The achievable data rate for user $(\ell,k)$ can be computed as~\cite{goldsmith2005wireless}
\begin{align}
        R_{\ell k}=\log|\mathbf{I}+\bfV_{\ell k}^H \bfH_{\ell k,\ell}^H \bfF^{-1}_{\ell k}\bfH_{\ell k,\ell}\bfV_{\ell k}|,
\end{align}
where
\begin{align}
        \bfF_{\ell k}=&\sum_{j=1,j\neq k}^{K}\bfH_{\ell k,\ell} \bfV_{\ell j}\bfV_{\ell j}^H \bfH^H_{\ell k,\ell}\nonumber\\
        &\qquad+\sum_{i=1,i\neq \ell}^{L}\sum_{j=1}^{K}\bfH_{\ell k,i} \bfV_{ij}\bfV_{ij}^H \bfH^H_{\ell k,i}+\sigma^2 \bfI_{N_r}.\label{equ:Flk}
\end{align}

We seek the optimal transmit beamformers $\underline{\bfV}$ to maximize the weighted sum rates:
\begin{subequations}
\label{prob}
\begin{align}
\max_{\underline{\bfV}}&\quad f_o(\underline{\bfV}):=\sum_{\ell =1}^{L}\sum_{k=1}^{K}w_{\ell k}R_{\ell k}\label{prob:MSR}\\
\text{s.t.}&\quad\sum_{k=1}^{K}\text{tr}(\bfV_{\ell k}\bfV^H_{\ell k})\le P_{\ell},\; \ell=1,2,\ldots,L,\label{cons:MSR}
\end{align}   
\end{subequations}
where the nonnegative weight $w_{\ell k}\ge 0$ reflects the priority of user $(\ell,k)$, and the constant $P_{\ell}$ is the power budget of BS $\ell$.

% \section{Model-Driven Approach}\label{model-Driven Methods}
\section{Existing Optimization-Based Methods}\label{model-Driven Methods}

\subsection{FP Method}

\begin{figure*}[htbp]
    \centering
    \begin{multline}\label{equ:MSE_n}
    f_n(\underline{\bfV},\underline{\bfGamma},\underline{\bfY},\underline{\bfZ}) = 
        \sum_{\ell,k} \Big[ \operatorname{tr} \left( 2\Re \left\{ \mathbf{V}_{\ell k}^H \mathbf{\Lambda}_{\ell k} +\bfV_{\ell k}^H(\lambda_{\ell}\mathbf{I}_{N_t}-\bfL_{\ell})\bfZ_{\ell k}\right\}+\bfZ^{H}_{\ell k}(\bfL_{\ell}-\lambda_{\ell}\mathbf{I}_{N_t})\bfZ_{\ell k}-\lambda_{\ell}\bfV_{\ell k}^H\bfV_{\ell k}\right)\\
        - \operatorname{tr} \left(\omega_{\ell k}\sigma^2(\mathbf{I}_{d}+\bfGamma_{\ell k})\bfY^H_{\ell k}\bfY_{\ell k}\right) + \omega_{\ell k} \log \left| \mathbf{I}_{d}+\bfGamma_{\ell k} \right| 
        - \operatorname{tr} \left( \omega_{\ell k} \mathbf{\bfGamma}_{\ell k} \right) \Big]\tag{15}
        \end{multline}
        \hrulefill
        % \vspace*{2pt}
\end{figure*}

\begin{algorithm}[t]
        \caption{FP for Multicell MIMO Beamforming}\label{alg:FP}
        \begin{algorithmic}[1]
        \STATE \textbf{input:} The current CSI.
        \STATE Initialize $\underline{\bfV}$ to feasible values under the power constraint.
        \REPEAT
            \STATE Update each $\bfY_{\ell k}$ by \eqref{opt_Ylk}.
            \STATE Update each $\bfGamma_{\ell k}$ by \eqref{opt_Gammalk}.
            \STATE Update each $\bfV_{\ell k}$ by \eqref{equ:Vlk_opt_fp}.
        \UNTIL{the objective value converges}
        \STATE \textbf{output:} Final beamforming matrix $\underline{\bfV}$
        \end{algorithmic}
\end{algorithm}

By the Lagrangian dual transform \cite{kaiming2018FP2}, the original objective $f_o(\underline{\bfV})$ is converted to
\begin{align}
        f_r(\underline{\bfV},\underline{\bfGamma})=\sum_{\ell =1}^{L}\sum_{k=1}^{K}w_{\ell k}\left[\log|\bfI_{d}+\bfGamma_{\ell k}|-\text{tr}(\bfGamma_{\ell k})\right.&\nonumber\\
        \left.+\text{tr}((\bfI+\bfGamma_{\ell k})\bfV_{\ell k}^H \bfH_{\ell k,\ell}^H \bfD^{-1}_{\ell k}\bfH_{\ell k,\ell}\bfV_{\ell k})\right],&\label{equ:MSR_r}
\end{align}
where
\begin{align}
        \bfD_{\ell k}&=\sum_{i=1}^{L}\sum_{j=1}^{K}\bfH_{\ell k,i} \bfV_{ij}\bfV_{ij}^H \bfH^H_{\ell k,i}+\sigma^2 \bfI_{N_r}.
\end{align}
The FP method then applies the quadratic transform \cite{kaiming2018FPI} to further recast $f_o(\underline{\bfV})$ into $f_{q}(\underline{\bfV},\underline{\bfGamma},\underline{\bfY})$ displayed in \eqref{equ:MSE_q} with
\setcounter{equation}{7}
\begin{align}
        \mathbf{\Lambda}_{\ell k}=w_{\ell k}\bfH^H_{\ell k,\ell}\bfY_{\ell k}(\mathbf{I}_{d}+\bfGamma_{\ell k}).\label{equ:Lambdalk}
\end{align}
The new objective $f_q(\underline{\bfV},\underline{\bfGamma},\underline{\bfY})$ is separately concave in $\underline{\bfV},\underline{\bfGamma},\underline{\bfY}$, so the FP algorithm allows iteratively optimizing these variables as
\begin{align}
        \bfY_{\ell k}&=\bfD_{\ell k}^{-1}\bfH_{\ell k,\ell}\bfV_{\ell k},\label{opt_Ylk}\\
        \bfGamma_{\ell k}&=\bfV_{\ell k}^H \bfH_{\ell k,\ell}^H \bfF^{-1}_{\ell k}\bfH_{\ell k,\ell}\bfV_{\ell k},\label{opt_Gammalk}\\
        \bfV_{\ell k}&=(\eta_{\ell}\bfI_{N_t}+\bfL _{\ell})^{-1}\bfLambda_{\ell k},\label{equ:Vlk_opt_fp}
\end{align}
where 
\begin{align}
        \bfL_{\ell} =\sum_{i=1}^{L}\sum_{j=1}^{K}w_{ij}\bfH^H_{ij,\ell}\bfY_{ij}(\mathbf{I}_{d}+\bfGamma_{ij})\bfY_{ij}^{H}\bfH_{ij,\ell},\label{equ:hatL}
\end{align}
and the Lagrange multiplier $\eta_\ell$ in \eqref{equ:Vlk_opt_fp} for the power constraint is computed as
\begin{align}
        \eta_{\ell}=\min\left\{\eta\ge0: \sum_{k=1}^{K}\text{tr}(\bfV_{\ell k}\bfV^H_{\ell k})\le P_{\ell}\right\},\label{equ:consV}
\end{align}
as summarized in Algorithm \ref{alg:FP}.
Note that the WMMSE algorithm \cite{Shi2011WMMSE,Christensen2008WMMSE} is a special case of Algorithm \ref{alg:FP}.

% \begin{remark}
% The above use of the FP technique recovers the WMMSE algorithm exactly. But we point out that there are other possible ways of using the FP technique, which lead to other iterative algorithms (and they may even outperform WMMSE in certain cases \cite{kaiming2018FP2}).
% \end{remark}

% \begin{remark}
% Similar to the WMMSE algorithm, the FP algorithm involves large matrix inversion and the bisection method for optimizing the Lagrange multipliers. Thus, although FP generalizes WMMSE, the above two drawbacks that prohibit deep unfolding still exist.
% \end{remark}

\subsection{FastFP Method \cite{Kaiming2024NFP,Zhang2023Enhangcing}}
The main drawback of the FP method is that it requires computing the large matrix inverse in \eqref{equ:Vlk_opt_fp}: recall that $\bfL_{\ell}$ is an $N_t\times N_t$ matrix and $N_t$ is a large number in the multicell MIMO setting. To eliminate the large matrix inversion, we can incorporate the following bound into the FP method:

\begin{Lemma}(Nonhomogeneous Bound\cite{Sunying2017MM})\label{lem:Nonhomogeneous_Bound}
Suppose that two Hermitian matrices $\bfL,\bfK\in \mathbb{H}^{m\times m}$ satisfy $\bfL\preceq \bfK$. Then for any two  matrices $\mathbf{X},\bfZ\in \mathbb{C}^{m\times m}$, one has
\begin{multline}
\label{iequ:NonhomoBound}
\operatorname{tr}(\mathbf{X}^H \mathbf{L} \mathbf{X}) \leq \operatorname{tr} \left( \mathbf{X}^H \mathbf{K} \mathbf{X} + 2\Re \{ \mathbf{X}^H (\mathbf{L} - \mathbf{K}) \mathbf{Z} \}\right.\\
\left.+ \mathbf{Z}^H (\mathbf{K} - \mathbf{L}) \mathbf{Z} \right),
\end{multline}
where the equality holds if $\bfZ=\bfX$.
\end{Lemma}

\begin{remark} 
\label{remark:lambda}
In a nutshell, we seek some matrix $\mathbf L$ in \eqref{iequ:NonhomoBound} that is easy to invert, so it is natural to let $\mathbf L = \lambda \mathbf I$. It remains to choose $\lambda\in\mathbb R$ to meet the condition $\bfL\preceq \bfK$. We can compute the largest eigenvalue of $\bfK$ and let $\lambda = \lambda_{\max}(\bfL)$. An alternative is to let $\lambda = \|\bfL\|_{F}$, but the gap between $\bfL$ and $\bfK$ becomes larger, so the convergence slows down.
\end{remark}

In light of Lemma \ref{lem:Nonhomogeneous_Bound} and Remark \ref{remark:lambda}, we further recast $ f_{q}(\underline{\bfV},\underline{\bfGamma},\underline{\bfY}) $ into $f_n(\underline{\bfV},\underline{\bfGamma},\underline{\bfY},\underline{\bfZ})$ as displayed in \eqref{equ:MSE_n}, where
\setcounter{equation}{15}
\begin{align}
        \lambda_\ell = \lambda_{\max}(\bfL_{\ell})\label{equ:lambda_compute}
\end{align}
When other variables are held fixed, each $\underline{\bfZ}$ in \eqref{equ:MSE_n} is optimally determined as
\begin{align}
        \bfZ_{\ell k}=\bfV_{\ell k}.\label{opt_Zlk}
\end{align}
When other variables are fixed, each $\bfV_{\ell k}$ for the current iteration $t$ is optimally determined based on that of the previous iteration $t-1$ as
\begin{align}
        \mathbf{V}^{(\tau)}_{\ell k} = \begin{cases}
                \widehat{\mathbf{V}}_{\ell k} & \text{if } \sum_{j=1}^K \|\widehat{\mathbf{V}}_{\ell j}\|_F^2 \leq P_{\ell} \\
                \sqrt{\frac{P_\ell}{\sum_{j=1}^K \|\widehat{\mathbf{V}}_{\ell j}\|_F^2}} \widehat{\mathbf{V}}_{\ell k} & \text{otherwise},
        \end{cases}\label{opt_Vlk}
\end{align}
where
\begin{align}
        \widehat{\mathbf{V}}_{\ell k} = \mathbf{V}^{(\tau-1)}_{\ell k} + \frac{1}{\lambda_\ell} (\bfLambda_{\ell k} - \bfL_\ell \mathbf{V}^{(\tau-1)}_{\ell k}).\label{opt_Vlk_hat}
\end{align}
Here and throughout, we use the superscript $\tau$ or $\tau-1$ to index the iteration. The optimal updates of $ \bfY_{\ell k} $ and $ \bfGamma_{\ell k} $ are the same as in \eqref{opt_Ylk} and \eqref{opt_Gammalk}. Algorithm \ref{alg:FastFP} summarizes the FastFP method.

\begin{algorithm}[t]
        \caption{FastFP for Multicell MIMO Beamforming}\label{alg:FastFP}
        \begin{algorithmic}[1]
        \STATE \textbf{input:} The current CSI.
        \STATE Initialize $\underline{\bfV}$ to feasible values under the power constraint.
        \REPEAT
            \STATE Update each $\bfZ_{\ell k}$ by \eqref{opt_Zlk}.
            \STATE Update each $\bfY_{\ell k}$ by \eqref{opt_Ylk}.
            \STATE Update each $\bfGamma_{\ell k}$ by \eqref{opt_Gammalk}.
            \STATE Update each $\bfV_{\ell k}$ by \eqref{opt_Vlk}.
        \UNTIL{the objective value converges}
        \STATE \textbf{output:} Final beamforming matrix $\underline{\bfV}$
        \end{algorithmic}

\end{algorithm}

\begin{figure}[tbp]
        \centering
        \includegraphics[width=0.4\textwidth]{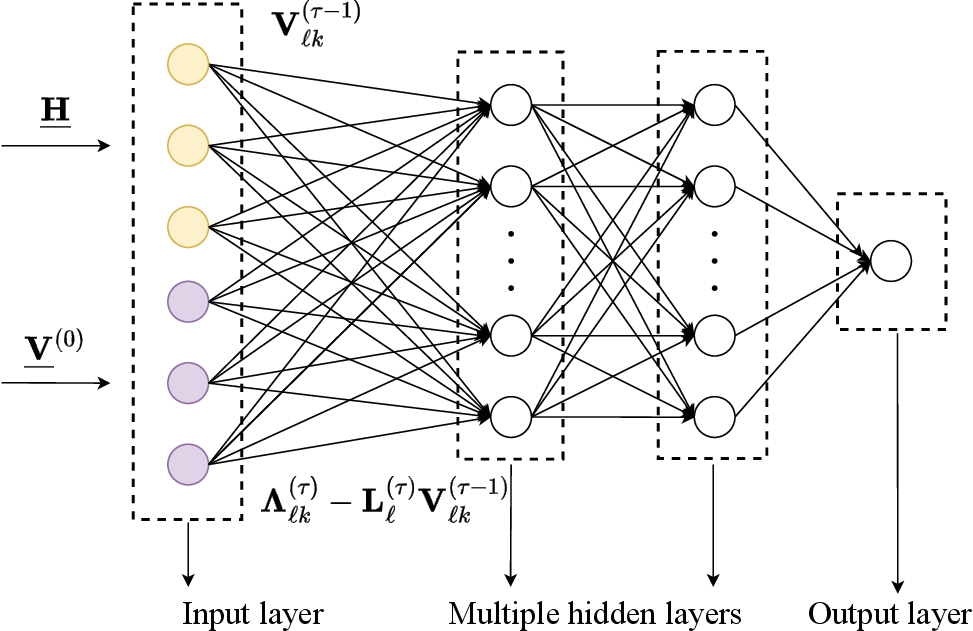}
        \caption{The DNN structure used in the DeepFP network. The DNN consists of one input layer, multiple hidden layers, and one output layer. The activation function in the hidden layers is the complex extension of ReLU.}
        \label{fig:DNN}
\end{figure}

\begin{figure*}[h]
        \centering
        \includegraphics[width=1.0\textwidth]{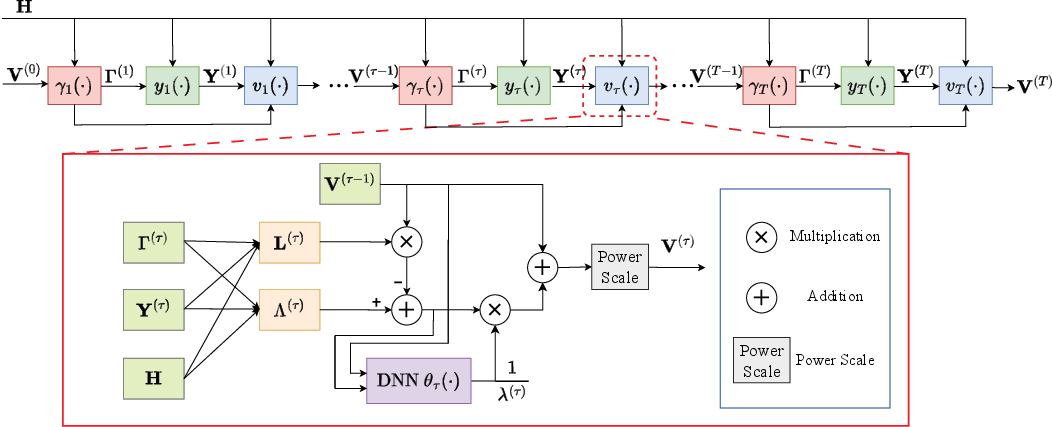}
        \caption{The architecture of our proposed DeepFP network. The modules \( z_\tau(\cdot), \gamma_\tau(\cdot), y_\tau(\cdot), v_\tau(\cdot) \) are designed based on \eqref{opt_Zlk}, \eqref{opt_Gammalk}, \eqref{opt_Ylk}, and \eqref{opt_Vlk}, respectively. In the module \( v_\tau(\cdot) \), the parameter \( \lambda^{(\tau)} \) is provided by the DNN \( \theta_{\tau} \), \( \bfL^{(\tau)} \) is determined by \eqref{equ:hatL}, and \( \bfLambda^{(\tau)} \) is determined by \eqref{equ:Lambdalk}. The DNNs in different layers of the DeepFP network have the same structure but do not share parameters.}
        \label{fig:unfolding}
\end{figure*}

% \section{Data-Driven Approach}\label{sec:unfolding}
\section{Proposed DeepFP for Multicell MIMO Beamforming}\label{sec:unfolding}

Recall that the FastFP method requires computing the largest eigenvalue of $\bfL$ to decide each $\lambda_\ell$, thus incurring a cubic computational complexity. This section introduces a deep unfolding method, called the DeepFP, that chooses $\lambda_\ell$ without explicit eigencomputation.

\subsection{Deep Unfolding for Iterative Optimization}

A generic iterative algorithm can be written in the following standard form as \cite{Hu2021IAIDNN}
\begin{align}
\underline{\mathbf{x}}^{(\tau)}=f_{\tau}(\underline{\mathbf{x}}^{(\tau-1)};\underline{\phi}),\label{iera_x}
\end{align}
where $\tau = 1, 2, \dots, T$ denotes the iteration index, $T$ is the total iteration number, $\underline{\mathbf{x}}$ is the optimization variable, the status variable $\underline{\phi}$ is a random variable that characterizes the uncertainty in the optimization problem (e.g., it is the random channel fading of the MIMO beamforming problem), and the $f_{\tau}$ is the iterate function that yields the new solution $\underline{\mathbf{x}}^{(\tau-1)}$ given the previous solution $\underline{\mathbf{x}}^{(\tau)}$ conditioned on the current status $\underline{\phi}$.

Deep Unfolding aims to unroll the iterative algorithm into a multi-layer sequential process. With a set of trainable parameters $\theta$, the deep unfolding method represents \eqref{iera_x} as a DNN layer:
\begin{align}
\underline{\mathbf{x}}^{(\tau)}=\mathcal{F}_{\tau}(\underline{\mathbf{x}}^{(\tau-1)};\theta_{\tau},\underline{\phi}),\label{iera_x_du}
\end{align}
where $\tau=1,2,...,T$ is reused to denote the layer index, $\mathcal{F}_{\tau}$ denotes the structure of deep unfolding network in the $\tau$th layer, and $\underline{\mathbf{x}}^{(\tau-1)}$ and $\underline{\mathbf{x}}^{(\tau)}$ are the input and output of the $\tau$th layer, respectively. In principle, after $\theta_\tau$ has been trained properly, $\mathcal{F}_{\tau}(\underline{\mathbf{x}}^{(\tau-1)};\theta_{\tau},\underline{\phi})$ is expected to behave similarly to $f_{\tau}(\underline{\mathbf{x}}^{(\tau-1)};\underline{\phi})$ for any possible $\underline{\phi}$.

\subsection{Optimizing $\lambda_\ell$ via DNN}

By specializing the above deep unfolding framework to the beamforming problem \eqref{prob} and the FastFP algorithm, we have the following correspondence:
\begin{align}
        &\underline{\mathbf{x}}= \{\bfGamma_{\ell k},\bfY_{\ell k},\bfV_{\ell k}\},\\
        &\underline{\phi}= \{\bfH_{\ell k,j},w_{\ell k},P_{\ell},\sigma^2\}.
\end{align}

Equation~\eqref{opt_Vlk} implies that the update of $\mathbf{V}_{\ell k}$ in FastFP follows a gradient projection form, where the scalar $1/{\lambda_{\ell}}$ serves as the step size for the update of all users in cell $\ell$.

We treat $\lambda_{\ell k}^{(\tau)}$ as a function of $\mathbf{V}_{\ell k}^{(\tau-1)}$ and the term $\bfLambda_{\ell k}^{(\tau)} - \bfL_\ell^{(\tau)} \mathbf{V}_{\ell k}^{(\tau-1)}$. Let $\theta_\tau(\cdot)$ denote the $\tau$th DNN layer in the unfolding network. The value of $\lambda_{\ell k}^{(\tau)}$ is then given by
\begin{align}
    \lambda^{(\tau)}_{\ell k} &= \theta_\tau(\mathbf{V}_{\ell k}^{(\tau-1)}, \bfLambda_{\ell k}^{(\tau)} - \bfL_\ell^{(\tau)} \mathbf{V}_{\ell k}^{(\tau-1)}). \label{equ:NN_lambda}
\end{align}
As \eqref{equ:NN_lambda} indicates, we think of $\lambda_{\ell k}^{(\tau)}$ as a function of $\mathbf{V}_{\ell k}^{(\tau-1)}$ and $\bfLambda^{(\tau)}_{\ell k} - \bfL^{(\tau)}_\ell \mathbf{V}_{\ell k}^{(\tau-1)}$. Here is the rationale of the above setting: in the FastFP algorithm, the beamforming matrix $\bfV^{(\tau)}_{\ell k}$ is updated as a linear combination of its value from the previous iteration and a new direction matrix $\bfLambda^{(\tau)}_{\ell k} - \bfL^{(\tau)}_\ell \mathbf{V}^{(\tau-1)}_{\ell k}$. When the number of iterations is small, $\bfV^{(\tau)}_{\ell k}$ significantly deviates from the new direction. Thus, $\lambda_{\ell k}^{(\tau)}$ should be large to accelerate convergence. As the number of iterations increases, $\bfV^{(\tau)}_{\ell k}$ approaches the stationary point, and $\bfLambda^{(\tau)}_{\ell k} - \bfL^{(\tau)}_\ell \mathbf{V}^{(\tau-1)}_{\ell k}$ approaches the zero vector. In this case, $\lambda_{\ell k}^{(\tau)}$ should be small to avoid oscillations. Thus, it leads to modeling $\lambda_{\ell k}^{(\tau)}$ as a function of $\bfV_{\ell k}^{(\tau-1)}$ and $\bfLambda_{\ell k}^{(\tau)} - \bfL^{(\tau)}_\ell \mathbf{V}^{(\tau-1)}_{\ell k}$.

In the FastFP algorithm, $\lambda_{\ell k}^{(\tau)}$ is set to the largest eigenvalue of $\bfL_\ell$ to ensure convergence. In contrast, the proposed DeepFP network need not require $\lambda_{\ell k}^{(\tau)}$ to satisfy \eqref{iequ:NonhomoBound}. Rather, our goal is seek a desirable $\lambda_{\ell k}^{(\tau)}$ through the DNN, to yield a better $\bfV^{(\tau)}_{\ell k}$. This goal can be achieved by choosing a smaller $\lambda_{\ell k}^{(\tau)}$ than that in \eqref{equ:lambda_compute}. According to Majorization-minimization (MM)~\cite{Sunying2017MM} theory, the WSR can be improved by optimizing its lower bound, i.e., the surrogate function $f_n(\underline{\bfV}, \underline{\bfGamma}, \underline{\bfY}, \underline{\bfZ})$. A smaller $\lambda_{\ell k}^{(\tau)}$ may result in a tighter lower bound, thereby accelerating the iterative process.

The DNN structure used in the DeepFP network consists of one input layer, multiple hidden layers, and one output layer, as shown in Fig.~\ref{fig:DNN}. The input to the DNN is the flattened $\bfV^{(\tau-1)}_{\ell k}$ and $\bfLambda^{(\tau)}_{\ell k} - \bfL^{(\tau)}_\ell \mathbf{V}^{(\tau-1)}_{\ell k}$. Instead of dealing with the real and imaginary parts separately, we directly use flattened complex matrices as the integrated input to the DNN. To achieve this, we extend the Rectified Linear Unit (ReLU) \cite{ReLU} activation function to support complex-valued data in the hidden layers. Specifically, the complex ReLU is defined as:
\begin{align}  
    \text{ReLU}_{\text{Complex}}(a + bi) = \max(a, 0) + \max(b, 0)i, 
\end{align}
where $i$ is the imaginary unit. We use $\Re(\cdot)$ to denote the activation function in the output layer to ensure that the output of the DNN is a real number.

\begin{algorithm}[t]
    \caption{DeepFP for Multicell MIMO Beamforming}\label{alg:DeepFP_TrainInfer}
    \begin{algorithmic}[1]
        \STATE \textbf{\textit{--Training Session--}}
        \STATE \textbf{input:} Randomly generated channel samples.
        \STATE \textit{Stage 1: Supervised Learning}
        
        \STATE Run Algorithm \ref{alg:FastFP} to obtain the solution $\underline{\bfV}^*$.
        \STATE Use $\underline{\bfV}^*$ as labels to train the DNN based on the loss function in \eqref{loss1}.

        \STATE \textit{Stage 2: Unsupervised Learning}
        \STATE Fine-tune the DNN parameters based on the loss function in \eqref{loss2}.
        
        \STATE \textbf{output:} The optimized DNN parameters.
        
        \vspace{1em}
        \STATE \textbf{\textit{--Test Session--}}
        \STATE \textbf{input:} The current CSI.
        \STATE initialize $\underline{\bfV}^{(0)}$ to feasible values under the power constraint.
        \FOR{$\tau = 1$ to $T$}
            \STATE $\underline{\boldsymbol{\Gamma}}^{(\tau)} \gets \gamma_\tau(\underline{\bfV}^{(\tau-1)};\underline{\phi})$
            \STATE $\underline{\mathbf{Y}}^{(\tau)} \gets y_\tau(\underline{\bfV}^{(\tau-1)};\underline{\phi})$
            \STATE $\lambda_{\ell k}^{(\tau)} \gets \theta_\tau(\bfV^{(\tau-1)}_{\ell k},\bfLambda^{(\tau)}_{\ell k} - \bfL^{(\tau)}_\ell \mathbf{V}^{(\tau-1)}_{\ell k})$
            \STATE $\underline{\mathbf{V}}^{(\tau)} \gets v_\tau(\underline{\bfV}^{(\tau-1)}, \underline{\bfGamma}^{(\tau)}, \underline{\bfY}^{(\tau)}; \underline{\lambda}^{(\tau)}, \underline{\phi})$
        \ENDFOR
        \STATE \textbf{output:} Beamforming matrices $\underline{\bfV}^{(T)}$.
    \end{algorithmic}
\end{algorithm}

\subsection{Unfolding Layers}

With the DNN $\theta_\tau(\cdot)$, the structure of the $\tau$th layer in the DeepFP network can be described as
\begin{align}
        \underline{\bfGamma}^{(\tau)} &= \gamma_\tau(\underline{\bfV}^{(\tau-1)};\underline{\phi}),\label{equ:NN_Gamma}\\
        \underline{\bfY}^{(\tau)} &= y_\tau(\underline{\bfV}^{(\tau-1)};\underline{\phi}),\label{equ:NN_Y}\\
        \lambda^{(\tau)}_{\ell k} &= \theta_\tau(\bfV^{(\tau-1)}_{\ell k},\bfLambda^{(\tau)}_{\ell k} - \bfL^{(\tau)}_\ell \mathbf{V}^{(\tau-1)}_{\ell k}),\\
        \underline{\bfV}^{(\tau)} &= v_\tau(\underline{\bfV}^{(\tau-1)}, \underline{\bfGamma}^{(\tau)}, \underline{\bfY}^{(\tau)}; \underline{\lambda}^{(\tau)}, \underline{\phi}),\label{equ:NN_v}
\end{align}
where  \eqref{equ:NN_Gamma}, \eqref{equ:NN_Y}, and \eqref{equ:NN_v} correspond to the iterative algorithm steps  \eqref{opt_Gammalk}, \eqref{opt_Ylk}, and \eqref{opt_Vlk}, respectively. Although the modules $\gamma_\tau(\cdot)$ and $y_\tau(\cdot)$ involve matrix inversion operations, but the matrices here are only $N_r\times N_r$, where $N_r$ is the number of receive antennas at each user terminal. Since this paper focus on a massive MIMO network, $N_r$ is typically much smaller than the number of transmit antennas $N_t$ at each base station. For this reason, we do not give much attention to the $N_r\times N_r$ matrix inversion, but only focus on eliminating the $N_t\times N_t$ matrix inversion in the module $v_\tau(\cdot)$.

But what if $N_r$ is also large? Actually, the elimination of the $N_r\times N_r$ matrix inversion is conceptually not different from that of the $N_t\times N_t$. As discussed in \cite{Kaiming2024NFP}, we simply need to further apply the nonhomogeneous bound in Lemma \ref{lem:Nonhomogeneous_Bound} to the update of $\bfY_{\ell k}$ and $\bf\Gamma_{\ell k}$, although the math notation would be much more complicated.

The full structure of the DeepFP network is depicted in Fig.~\ref{fig:unfolding}. The variables $\underline{\bfL}^{(\tau)}$ and $\underline{\bm\Lambda}^{(\tau)}$ are computed based on \eqref{equ:hatL} and \eqref{equ:Lambdalk}, respectively. The DNNs across different layers of the unfolding network are based on the same structure (e.g., the number of hidden layers, the number of neurons per layer, and the activation functions). The module named "Power Scale" represents scaling beamforming vectors to satisfy the power constraints. This block corresponds to \eqref{opt_Vlk}. It aims to enforce the transmit power constraint for the beamforming matrix produced by the DNN.

Actually, the core question our paper aims to answer is whether it is worthwhile to stick to the MM properties. Indeed, the MM properties can warrant convergence, but at the cost of computation---we have to repeatedly compute the inverse of a large $N_t\times N_t$ matrix. The nonhomogeneous FP can eliminate the matrix inverse but then $\lambda_\ell$ is difficult to decide. It is shown in \cite{Kaiming2024NFP} that $\lambda_\ell=\lambda_{\max}(L_\ell)$ can guarantee the MM properties when the same $\lambda_\ell$ is used for all the users in cell $\ell$, but then the performance hurts because it sacrifices the flexibility in choosing $\lambda_\ell$. In contrast, this paper relaxes the MM constraint to allow $\lambda_\ell$ to be separately tuned for each individual user.

\subsection{Training Strategy}

We adopt a hybrid training strategy that comprises two stages. In the first stage, for a given channel sample $\underline{\bfH}$, let $\underline{\bfV}^*$ denote the solution obtained from the FastFP algorithm (Algorithm \ref{alg:FastFP}), and let $\underline{\bfV}^{(T)}$ denote the output of the unfolding network at the final layer. The first training stage employs supervised learning, with $\underline{\bfV}^*$ serving as the label with respect to the sample $\underline{\bfH}$. The MSE between $\underline{\bfV}^*$ and $\underline{\bfV}^{(T)}$ is adopted as the loss function in the first stage:
\begin{align}
    \mathrm{LOSS}_1 = \frac{1}{KL} \sum_{\ell=1}^{L} \sum_{k=1}^{K} \|\bfV_{\ell k}^{(T)} - \bfV_{\ell k}^*\|_2^2. \label{loss1}
\end{align}
In the second stage, we switch to the unsupervised learning, using the WSR function of $\underline{\bfV}^{(T)}$ as the loss function, that is
\begin{align}
    \mathrm{LOSS}_2 = -\frac{1}{KL}\sum_{\ell=1}^{L} \sum_{k=1}^{K} w_{\ell k} R_{\ell k}. \label{loss2}
\end{align} 
Although fully unsupervised learning has been shown to be feasible \cite{hojatian2021unsupervised}, we employs this hybrid training strategy that combines supervised and unsupervised learning.

% Specifically, in the first stage, the DNN is trained with the model-driven FastFP-generated solutions as labels to ensure rapid convergence and robust initialization. Subsequently, in the second stage, the DNN is fine-tuned using the actual weighted sum-rate objective as the loss function.  Fig.~\ref{fig:exp_train_strategies} compares this two-stage hybrid training strategy with the fully supervised training and the fully unsupervised training scheme. It is observed that the hybrid training strategy strikes much better trade-off between the convergence speed and the achieved performance than the benchmarks.}

Regarding the parameter initialization, each $\underline{\bfV}^{(0)}$ is randomly and independently generated according to the standard complex Gaussian distribution $\mathcal{C} \mathcal{N}(0,1)$, followed by a scaling process to meet the power constraint $\sum_{k=1}^{K}\operatorname{tr}(\bfV_{\ell k}\bfV_{\ell k}^H) = P_{\ell}$ for each BS. Moreover, for the supervised learning at the first stage, the FastFP algorithm and the unfolding network use the same starting point $\underline{\bfV}^{(0)}$. Algorithm~\ref{alg:DeepFP_TrainInfer} summarizes the training and inference procedures of the DeepFP method.

During the training stage, we use a large set of randomly generated CSI drawn from the same distribution to train the DNN. After training, the DNN is fixed and can be thought of as a deterministic function that takes the specific CSI values as input and yields the corresponding output. In other words, the DNN itself depends on the channel distribution rather than the specific CSI values. Thus, there is no need to retrain the DNN when the beamforming problem is considered for a new realization of CSI from the same distribution.

The current DeepFP requires centralized training, but there are several ways to make it distributed. One possible method is to group the cells into clusters and then train the DNN on a per-cluster basis. Another possible method is to incorporate federated learning into the training stage.

\begin{figure}[!t]
        \begin{minipage}{1\linewidth}
            \centering
            \subfloat[Different choices of learning rate.]{\includegraphics[width=0.9\linewidth]{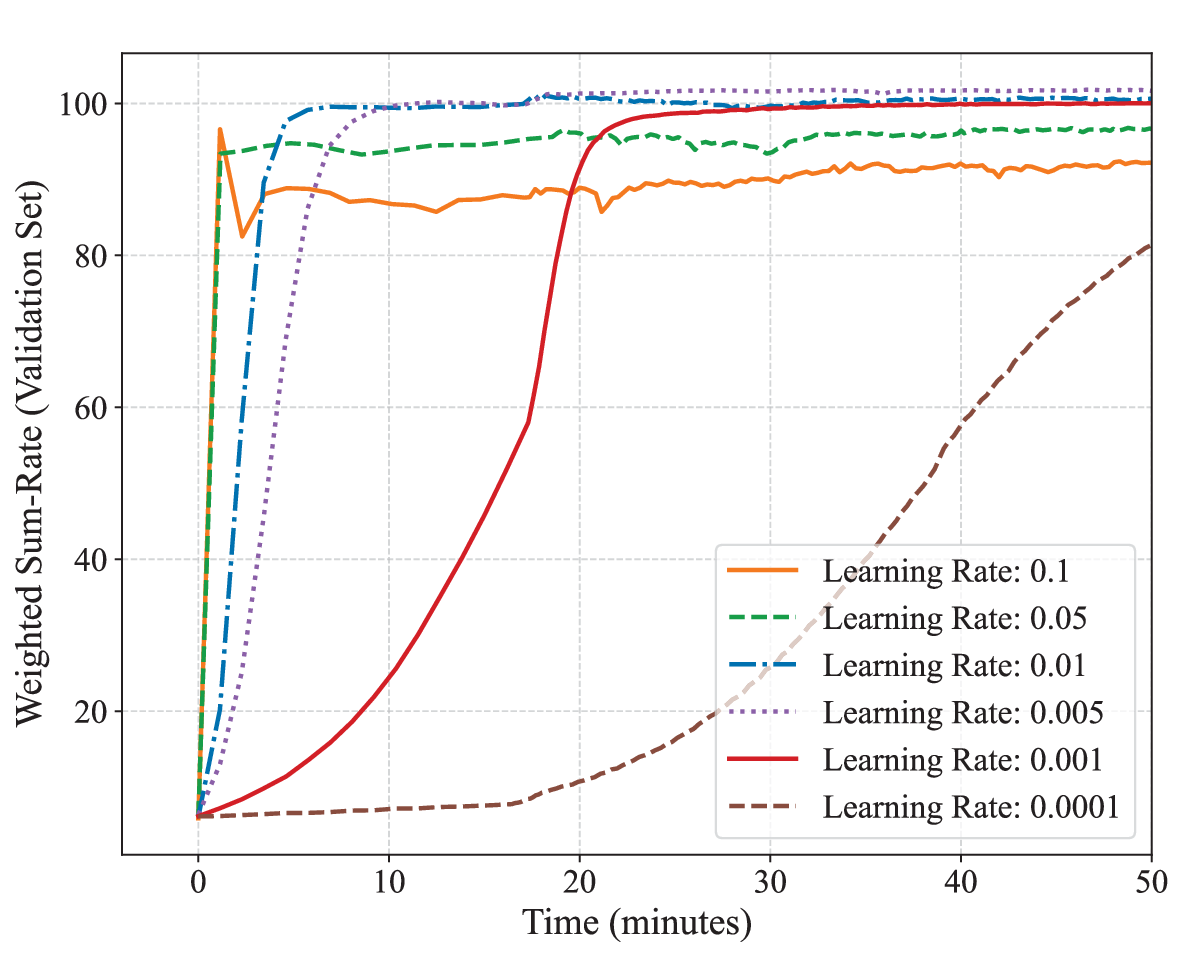}}
    
            \subfloat[Different choices of batch size.]{\includegraphics[width=0.9\linewidth]{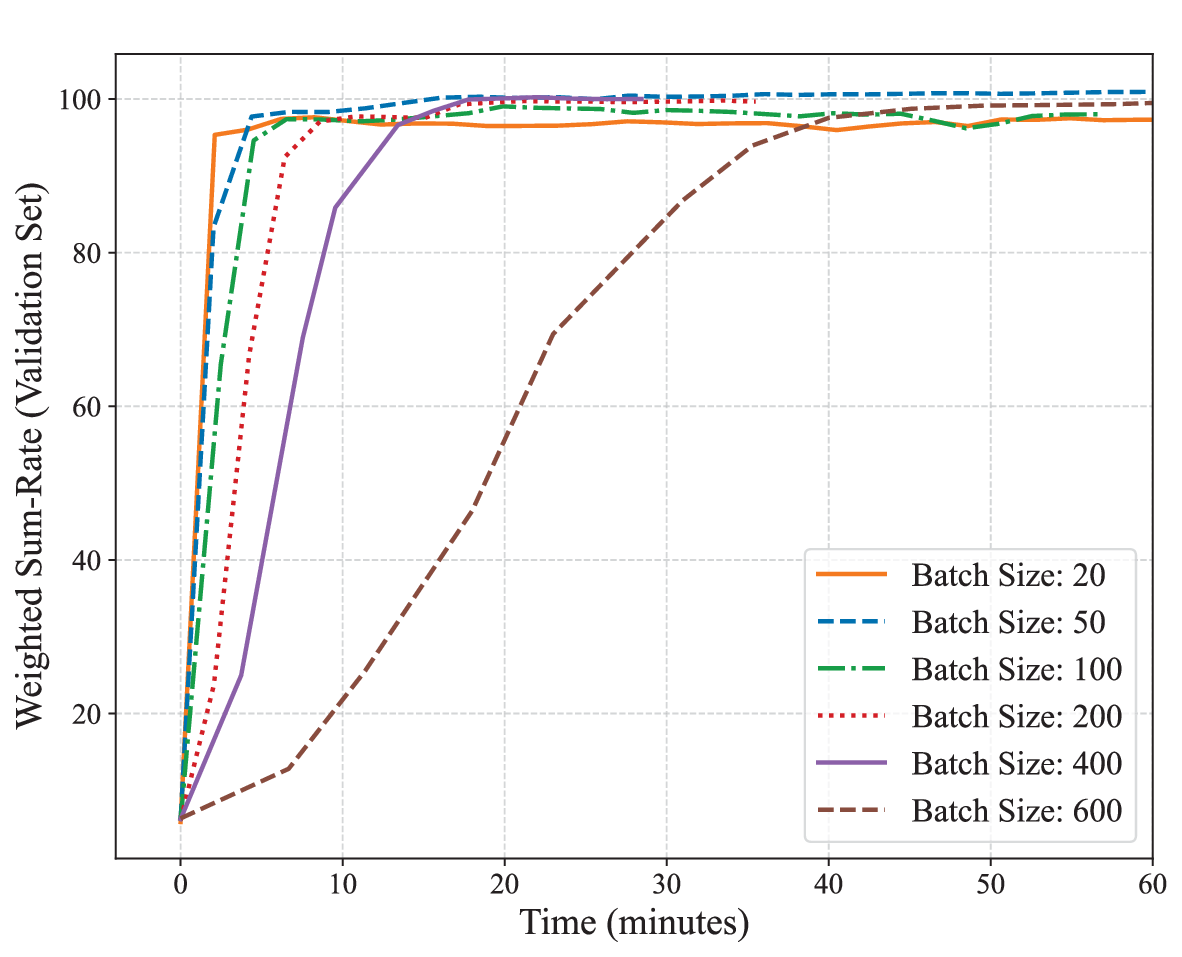}}
        \end{minipage}
        \caption{The WSR performance on validation dataset during training process for different learning rate (a) and batch size (b).}\label{fig:exp1}
\end{figure}

\subsection{Complexity Analysis}

Consider an $L$-cell MIMO system where each cell is equipped with $N_t$ transmit antennas and serves $K$ users. Each user is equipped with $N_r$ receive antennas. The number of data streams is $d$. The computational complexity of Algorithm~\ref{alg:FastFP} at each iteration is given by
\begin{multline}
\mathcal{O}\left(LN_t^3+L^2K^2(N_t+N_r)N_rd\right.\\
\left.+LKN_r^2(N_r+d)+LK^2(N_t+d)d^2\right),\label{equ:Fastfp_flops}
\end{multline}
where the term $\mathcal{O}(LN_t^3)$ is for the computation of the largest eigenvalue in \eqref{equ:lambda_compute}.

In the DeepFP network, the largest eigenvalue computation is replaced by the forward propagation of the DNN, with all other operations remaining unchanged. Consequently, the computational complexity for each layer is
\begin{align}
&\mathcal{O}\left(LKU(2N_td+(M_{\text{hid}}-1)U+1)+L^2K^2(N_t+N_r)N_rd\right.\notag\\
&\left.+LKN_r^2(N_r+d)+LK^2(N_t+d)d^2\right),\label{equ:deepfp_flops}
\end{align}
where $M_{\text{hid}}$ represents the number of hidden layers in the DNN, and $U$ represents the number of neural units in each hidden layer. Compared to the FastFP algorithm, the DeepFP network achieves lower computational complexity. It is worth emphasizing that the number of transmit antennas $N_t$ at the base station is the dominant factor, especially considering a massive MIMO network. With respect to $N_t$, \eqref{equ:Fastfp_flops} shows that the eigenvalue-based benchmark method leads to a complexity of $O(N^3_t)$, while \eqref{equ:deepfp_flops} shows that our method DeepFP leads to a much lower complexity of $O(N_t)$.

\begin{figure}[tbp]
    \centering
    \includegraphics[width=0.45\textwidth]{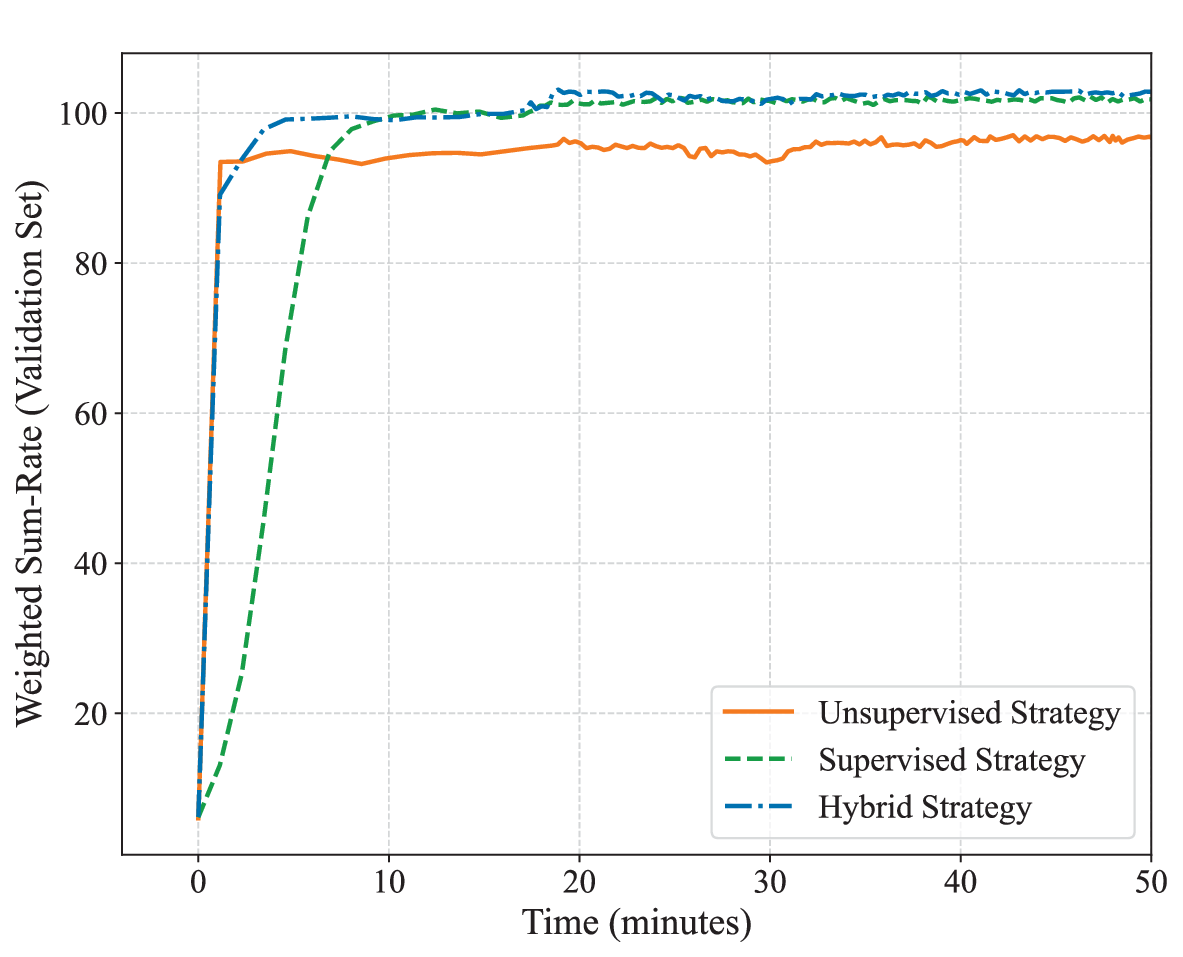}
    \caption{Validation performance comparison of three training strategies: Unsupervised, Supervised, and Hybrid. The curves show the WSR over training time, with each strategy exhibiting distinct convergence behavior.}
    \label{fig:exp_train_strategies}
\end{figure}

\begin{figure}[tbp]
        \centering
        \includegraphics[width=0.45\textwidth]{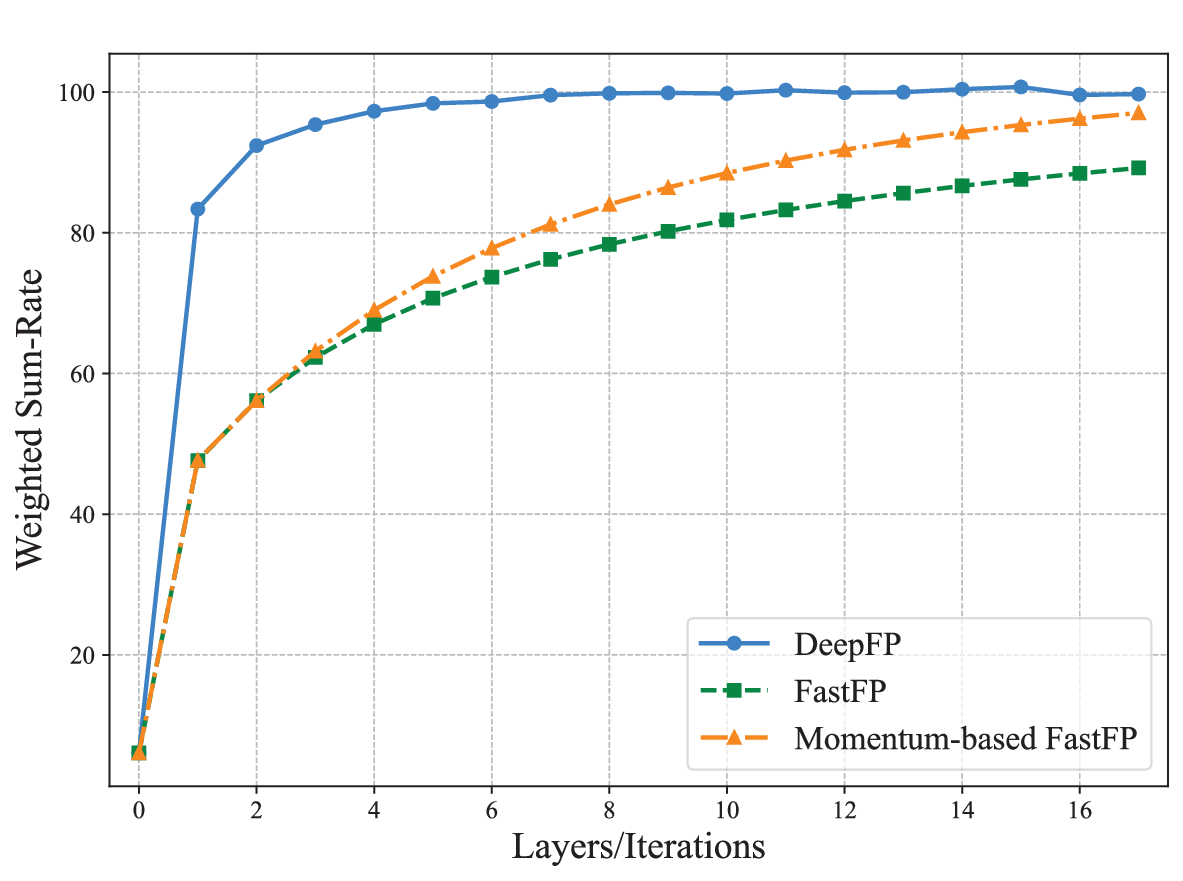}
        \caption{Weighted sum-rate performance of the DeepFP network, the FastFP algorithm, and the momentum-based FastFP algorithm. For FastFP and momentum-based FastFP, the curves depict the WSR after $i$ iterations. For DeepFP, the curve shows the WSR achieved by a network with $i$ layers.}
        \label{fig:exp_layers}
\end{figure}

\begin{figure}[tbp]
        \centering
        \includegraphics[width=0.45\textwidth]{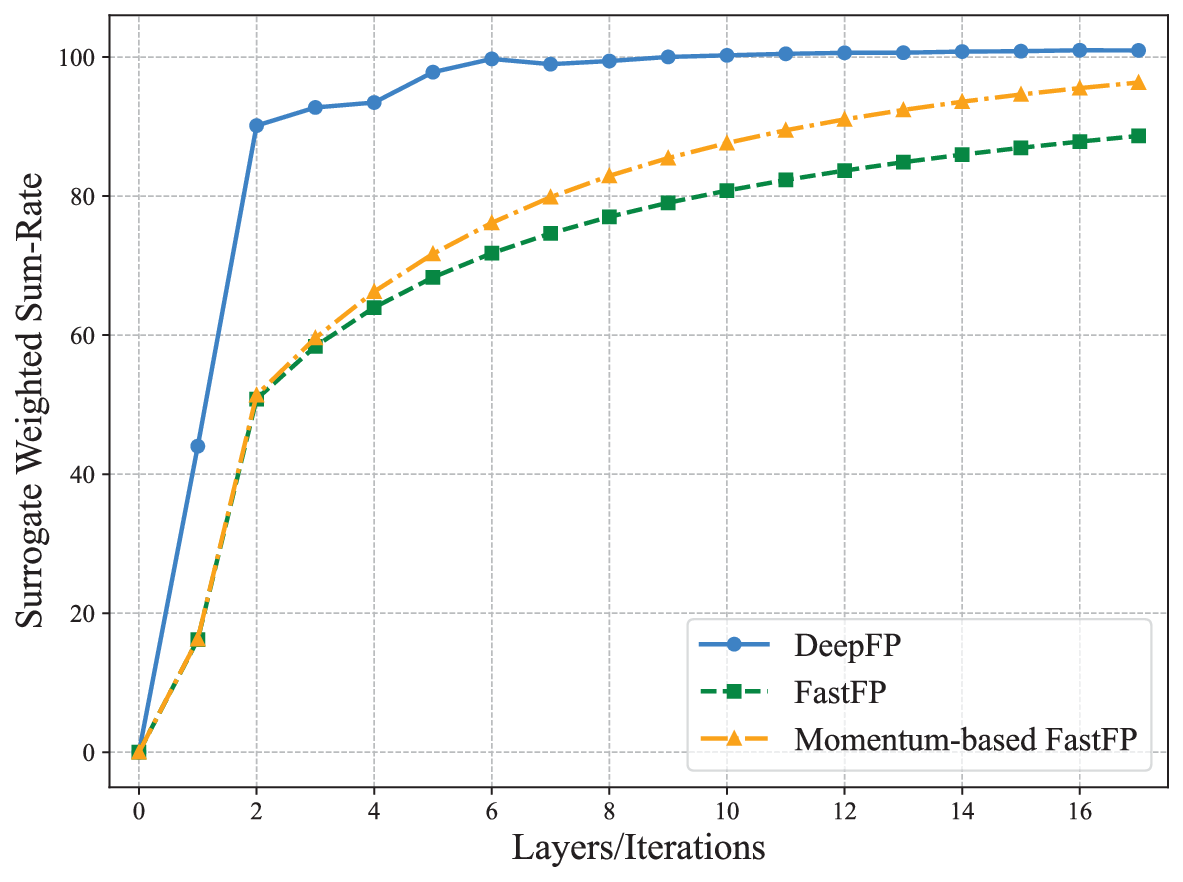}
        \caption{Surrogate weighted sum-rate performance of the DeepFP network and the FastFP algorithm.}
        \label{fig:surrogate_convergence}
\end{figure}

\begin{figure}[tbp]
        \centering
        \includegraphics[width=0.45\textwidth]{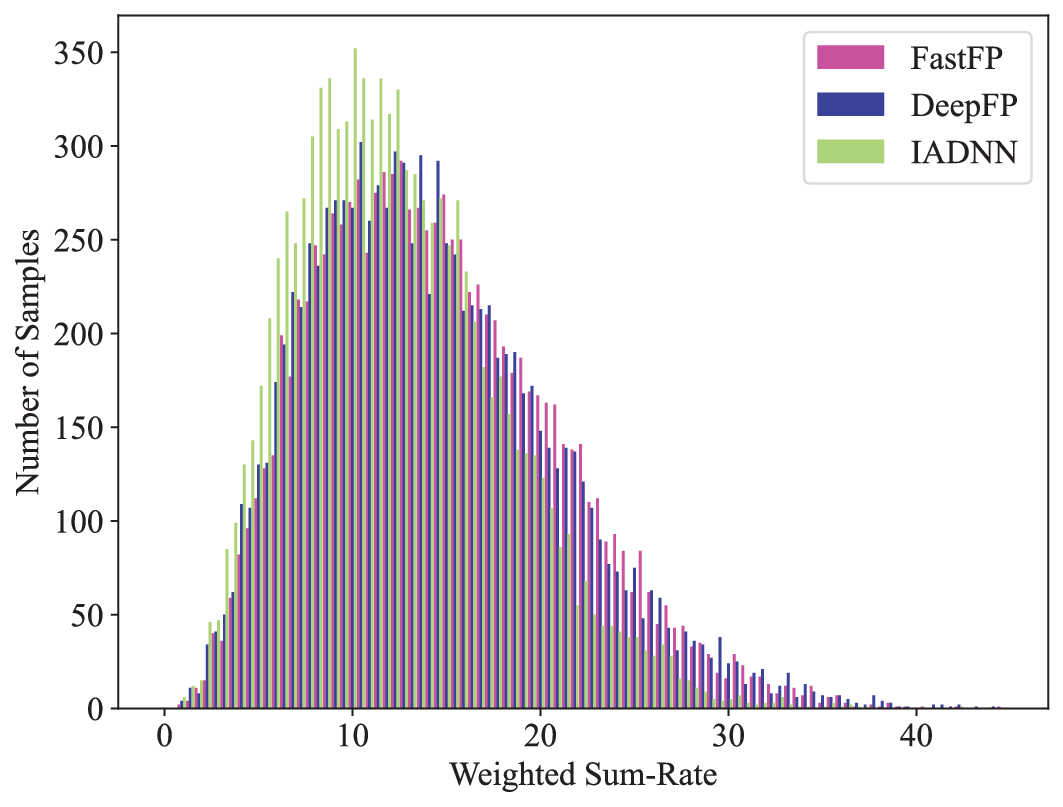}
        \caption{Distributions of the DeepFP network and baseline algorithms in single cell MIMO system with $N_t=64,N_r=4,d=2,K=6$.}
        \label{fig:exp_singlecell_hist}
\end{figure}

\begin{figure}[tbp]
        \centering
        \includegraphics[width=0.45\textwidth]{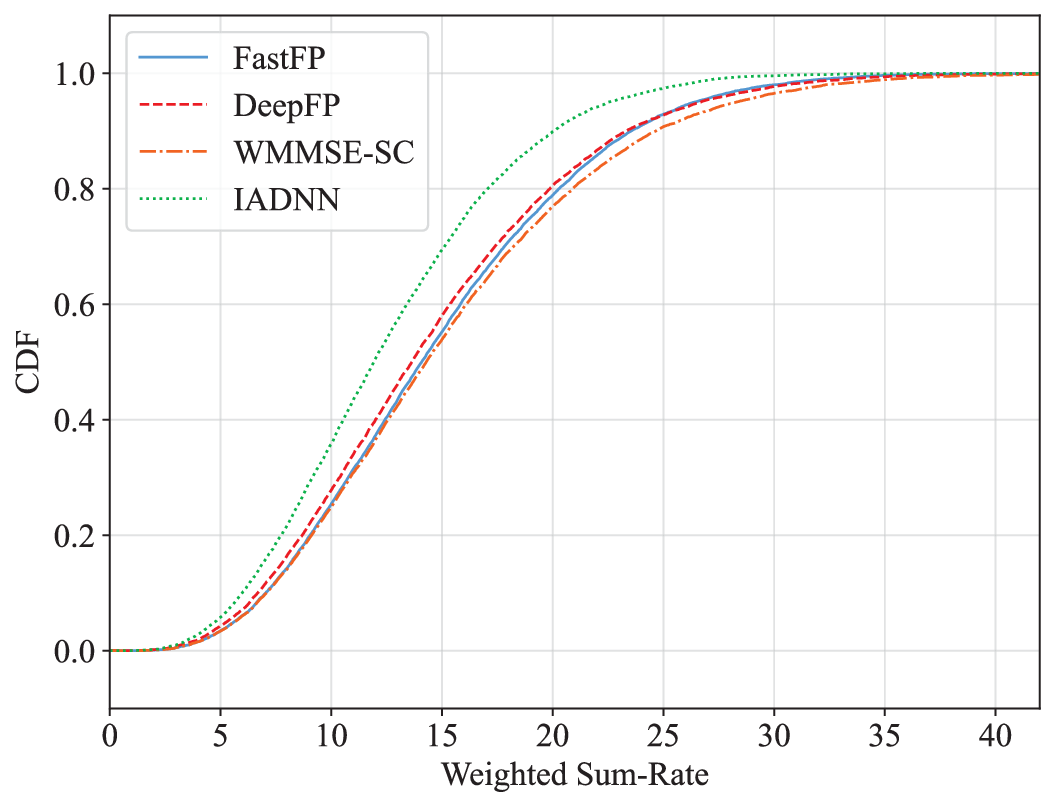}
        \caption{The CDF that describes the rates achieved by different algorithms in single cell MIMO system with $N_t=64,N_r=4,d=2,K=6$.}
        \label{fig:exp_singlecell_cdf}
\end{figure}

\begin{figure}[t]
        \centering
        \includegraphics[width=0.45\textwidth]{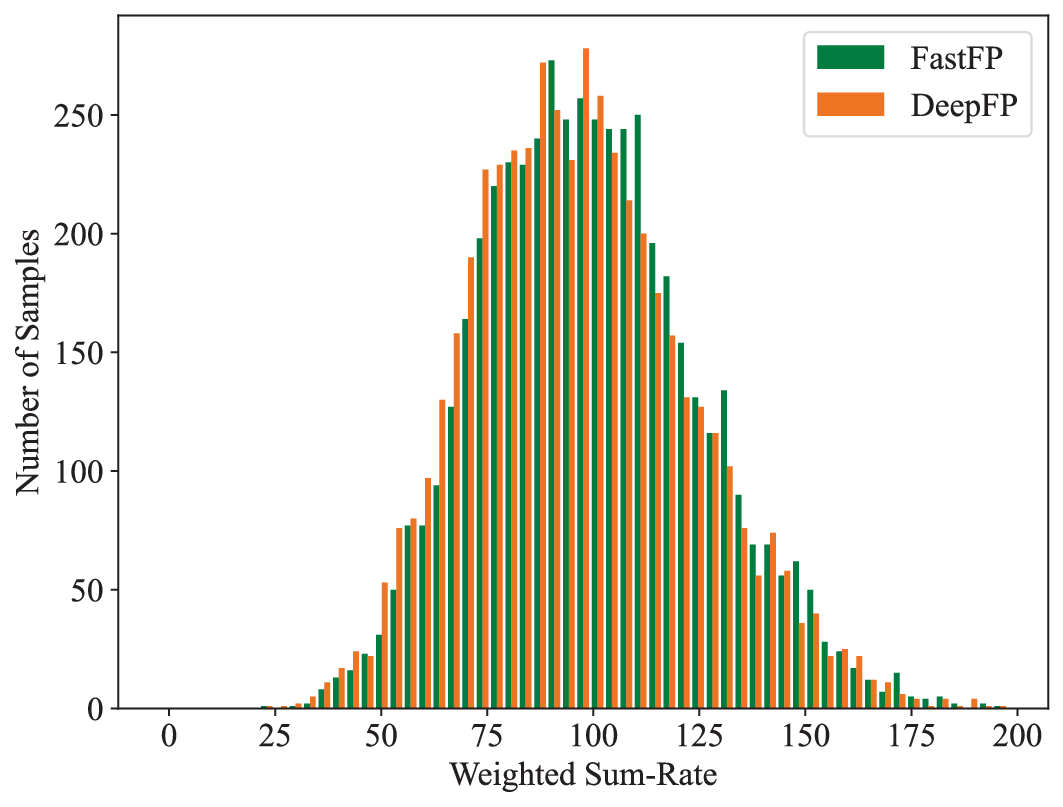}
        \caption{Distributions of the DeepFP network and the FastFP algorithm in 7-cell MIMO with $N_t=64,N_r=4,d=2,K=6$.}
        \label{fig:exp_multicell_hist}
\end{figure}

\begin{figure}[t]
        \centering
        \includegraphics[width=0.45\textwidth]{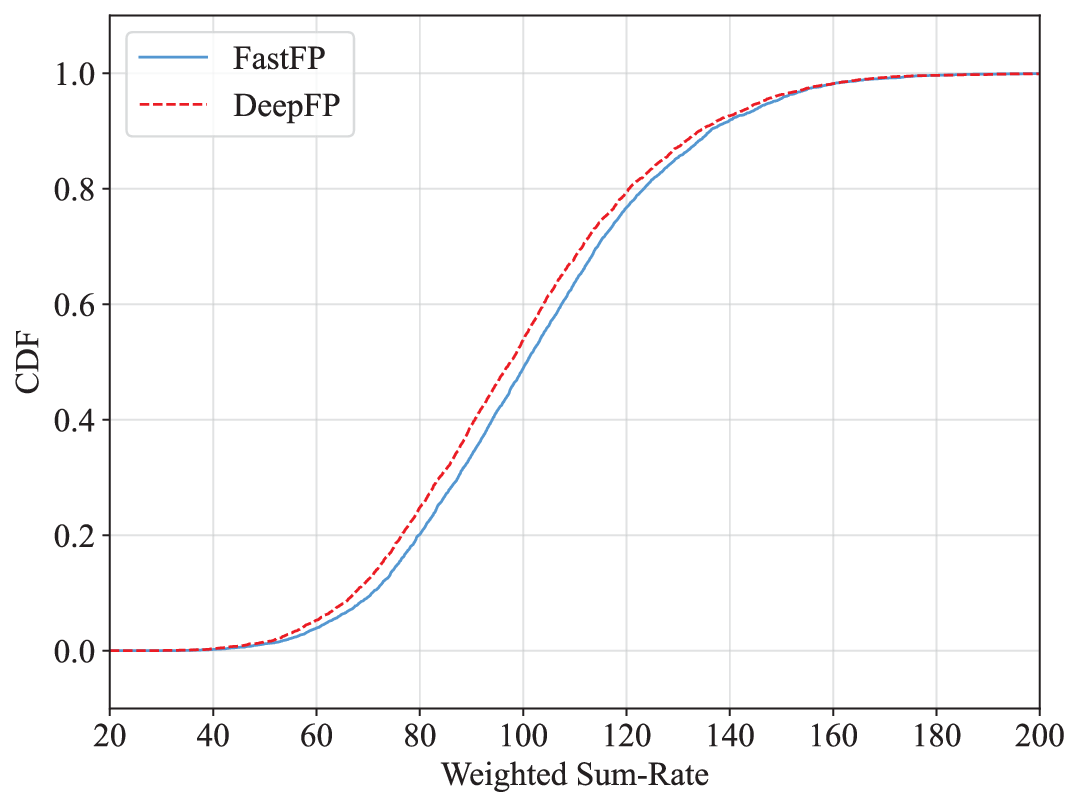}
        \caption{The CDF that describes the rates achieved by different algorithms in 7-cell MIMO with $N_t=64,N_r=4,d=2,K=6$.}
        \label{fig:exp_multicell_cdf}
\end{figure}

\section{Experimental Results}\label{sec:results}

In this section, we evaluate the performance of the proposed DeepFP network versus model-driven algorithms and existing unfolding algorithms. First, we evaluate how the different training strategies impact the optimization performance. Next, we try out a variety of wireless network examples. Finally, we validate the generalizability of the proposed DeepFP network by using different settings for training and test. The proposed DeepFP network is implemented in Python 3.10.0 with PyTorch 2.4.1. The system runs on a desktop with an Intel i7-13700 Central Processing Unit (CPU) clocked at 3.4 GHz and 64 GB of Random Access Memory (RAM). A Graphics Processing Unit (GPU) RTX 4080 is used during training to reduce training time, but not during testing. All algorithms, including DeepFP and the classical baselines, are implemented in PyTorch for consistency. For fair runtime comparison, all inference tests are conducted on CPU without GPU acceleration.

\begin{table}[t]
\renewcommand\arraystretch{1.2}
        \centering
        \caption{Weighted Sum-Rate and Computational Performance for Single-Cell MIMO with $N_t=64,N_r=4,d=2,K=6$.}\label{tab:singlecell}
        \begin{tabular}{|c|c|c|c|c|c|c|}
            \hline
                Algorithm & Weighted Sum-Rate & CPU Time (Sec.)\\
            \hline
            DeepFP & 14.664 (98.9\%)& 0.053 (14.0\%)\\
            FastFP & 14.826 (100.0\%)& 0.378 (100.0\%)\\
            FastFP (76 iterations) & 14.664 (98.9\%)& 0.287 (76.0\%)\\
            WMMSE-SC & 15.270 (103.0\%)& 0.563 (148.9\%)\\
            IADNN & 12.540 (84.9\%)& 0.055 (14.5\%)\\
            \hline
        \end{tabular}
\end{table}

\begin{table}[t]
\renewcommand\arraystretch{1.2}
        \centering
        \caption{Weighted Sum-Rate and Computational Performance for Multicell MIMO with $N_t=64,N_r=4,d=2,K=6$.}\label{tab:multicell}
        \begin{tabular}{|c|c|c|c|c|c|c|}
            \hline
                Algorithm & Weighted Sum-Rate & CPU Time (Sec.)\\
            \hline
            DeepFP & 99.474 (97.4\%)& 0.275 (11.7\%)\\
            FastFP & 102.186 (100.0\%)& 2.333 (100.0\%)\\
            FastFP (56 iterations) & 99.480 (97.4\%)& 1.306 (56.0\%)\\
            GCN-WMMSE &     91.011 (89.1\%)   &  0.604(25.9\%)       \\
            \hline
        \end{tabular}
\end{table}

\subsection{Experimental Setup}
\subsubsection{Dataset Generation}

We generate channel data from a 7-hexagonal-cell MIMO system as considered in~\cite{Kaiming2024NFP}. Within each cell, the BS is located at the center, and the $K$ downlink users are randomly distributed. Each BS and user are equipped with $N_t$ and $N_r$ antennas, respectively. The number of data streams is $d \leq N_r$. The weights of all users are set to be equal. The distance between adjacent BSs is $D = 0.8$ km. The maximum transmit power of each BS is $20$ dBm, and the background noise power is $-90$ dBm. The distance-dependent path loss of the downlink is modeled as $128.1 + 37.6 \log_{10} r + \xi$ (in dB), where $r$ denotes the distance from the BS to the user (in kilometers). $\xi$ is a zero-mean Gaussian random variable with an $8$ dB standard deviation to account for the shadowing effect.

% We randomly generate $50,000$ channel samples based on the above model, and divide them into training, validation, and test sets in a $0.70:0.15:0.15$ ratio. During training, the dataset is divided into multiple minibatches of the same batch size. The DeepFP network is trained over multiple epochs. The validation set is used to adjust the learning rate during training, while the test set is used to evaluate the performance of the trained network. We use the FastFP algorithm to generate the labels for the first training stage on the training set. Since FastFP avoids the use of bisection, it is well-suited for parallel execution, allowing the label generation process to be completed in just 40 minutes.

\subsubsection{Parameters Selection}

For all our numerical results, the DNN consists of two hidden layers, one input layer, and one output layer. Unless explicitly stated, each hidden layer contains 64 neurons. We first investigate the impact of batch size and learning rate on convergence performance. We set $N_t = 64$, $N_r = 4$, $K = 6$, and $d = 2$. The DeepFP network has $T = 8$ layers. Fig.\ref{fig:exp1} shows how the weighted sum-rate of the validation set changes during training for different batch size and learning rate settings. The results show that a larger learning rate speeds up convergence. However, an excessively large learning rate may cause instability and lower WSR performance. Thus, based on the results in Fig.\ref{fig:exp1}(a), we select an initial learning rate of 0.005, which is gradually decreased during the training process. The results in Fig.~\ref{fig:exp1}(b) show that as the batch size increases, the convergence rate initially improves and then decreases. This occurs because excessively large batch sizes result in longer processing times per minibatch due to memory limitations. Therefore, we choose a batch size of 200 to balance WSR performance and convergence rate. We further compare three training strategies: supervised, unsupervised, and hybrid. As shown in Fig.~\ref{fig:exp_train_strategies}, the mixed training scheme can strike a better trade-off between the convergence speed and the ultimate performance than the benchmarks.

We then analyze the effect of the number of unfolding layers $T$ in the DeepFP network on WSR performance. Networks with varying numbers of unfolding layers $T$ are trained, and their WSR performance is evaluated on the test set. The results in Fig.~\ref{fig:exp_layers} show that as $T$ increases, the WSR performance improves initially but begins to fluctuate once $T$ exceeds 8. Since the inference time of the DeepFP network grows linearly with $T$, we select $T = 8$ to balance WSR performance with inference time. Moreover, Fig.~\ref{fig:exp_layers} demonstrates the significant performance advantage of the DeepFP network compared to the FastFP algorithm and the momentum-based
FP. It shows that the DeepFP outperforms the benchmarks based on the MM properties. The DeepFP network achieves far superior performance compared to the FastFP algorithm when the number of layers in the DeepFP network equals the number of iterations in the FastFP algorithm. We further compare the surrogate function $f_n(\cdot)$ in \eqref{equ:MSE_n} of DeepFP, FastFP and the momentum-based FastFP in Fig.~\ref{fig:surrogate_convergence}. As shown, although the monotonic performance is no longer guaranteed by the DeepFP, the occasional performance drops are negligible overall, and the rate performance improves much more quickly thanks to the more aggressive choice of stepsizes.

\begin{table}[t]
\renewcommand\arraystretch{1.2}
        \centering
        \caption{Weighted Sum-Rate and Computational Performance of the DeepFP network for Multicell MIMO with $N_t=64,N_r=4,d=4$ for different $K$.}\label{tab:Kresults}
        \begin{tabular}{|c|c|c|c|c|c|c|}
            \hline
                $K$ & Weighted Sum-Rate & CPU Time (Sec.)& Iterations by FastFP \\
            \hline
            $6$ & 128.796 (92.8\%)& 0.285 (11.2\%)&23\\
            $9$ & 157.554 (90.3\%)& 0.584 (12.1\%)&21\\
            $15$ & 203.895 (86.5\%)& 1.678 (7.6\%)&19\\
            \hline
        \end{tabular}
\end{table}

\begin{table}[t]
\renewcommand\arraystretch{1.2}
        \centering
        \caption{Weighted Sum-Rate Performance of the DeepFP network with $N_t=64,N_r=4,K=6$ for different $d$. The DeepFP network is trained with $d=4$.}\label{tab:gene_d}
        \begin{tabular}{|c|c|c|c|c|c|c|}
            \hline
                $d$ & Weighted Sum-Rate  (bit/sec.)& Iterations by FastFP \\
            \hline
            $1$ & 66.486 (96.5\%)& 57\\
            $2$ & 97.074 (95.0\%)& 40\\
            $3$ & 115.596 (94.6\%)& 32\\
            $4$ & 128.796 (92.8\%)& 23\\
            \hline
        \end{tabular}
\end{table}

\subsection{WSR Maximization for Different Wireless Networks}

\subsubsection{Single-Cell Performance}

We evaluate the WSR performance of the DeepFP network under different network sizes. We begin with a single-cell MU-MIMO system with $N_t = 64$, $N_r = 4$, $K = 6$, and $d = 2$. The following three algorithms are selected as baseline algorithms: 
\begin{enumerate}
        \item \textbf{FastFP Algorithm}: The result of the FastFP Algorithm is taken as the output of Algorithm \ref{alg:FastFP} after $100$ iterations.
        \item \textbf{WMMSE-SC Algorithm}: The WMMSE-SC algorithm first uses WMMSE to solve a unconstrained WSR problem, and then scales the solution to satisfy the power constraints. This method avoids the bisection method but retains large matrix inversion, and it has theoretical guarantees only in the single-cell case. The result after 100 iterations is taken as the output of the WMMSE-SC algorithm.
        \item \textbf{IADNN}: The Iterative Algorithm-Induced Deep Unfolding Neural Network (IAIDNN) \cite{Hu2021IAIDNN} unfolds the WMMSE-SC algorithm for single-cell MIMO systems. IAIDNN eliminates large matrix inversions by introducing trainable matrices that approximate matrix inversion based on the first-order Taylor expansion. We implemented the original network structure proposed in \cite{Hu2021IAIDNN} using PyTorch, following the training settings recommended in \cite{Hu2021IAIDNN}. The number of layers in IAIDNN is set to $7$, as used in \cite{Hu2021IAIDNN}.
\end{enumerate}

We evaluate the WSR performance of the DeepFP network and baseline algorithms using the same test data. The average WSR and CPU time are computed from $10,000$ test samples, with the results presented in Table \ref{tab:singlecell}. We also report the results of FastFP after $76$ iterations, which achieves the same WSR performance as the DeepFP network. The WSR performance and runtime of each algorithm are compared to those of the FastFP algorithm, using percentages for clarity. The results show that the DeepFP network achieves $98.9\%$ of the WSR achieved by FastFP after $100$ iterations, while using only $14.0\%$ of its runtime. The FastFP algorithm requires $76$ iterations to achieve the same WSR performance as the DeepFP network, resulting in nearly five times the runtime. Moreover, our algorithm outperforms IADNN in WSR performance with a similar computation time.

The distribution and cumulative distribution function (CDF) of the WSR performance achieved by different algorithms are shown in Fig.\ref{fig:exp_singlecell_hist} and Fig.\ref{fig:exp_singlecell_cdf}, respectively. Each distribution and its corresponding CDF are based on results from $10,000$ test samples. The results indicate that the proposed DeepFP network closely matches the performance of the FastFP algorithm and outperforms the IADNN algorithm.

\begin{figure}[t]
        \centering
        \includegraphics[width=0.45\textwidth]{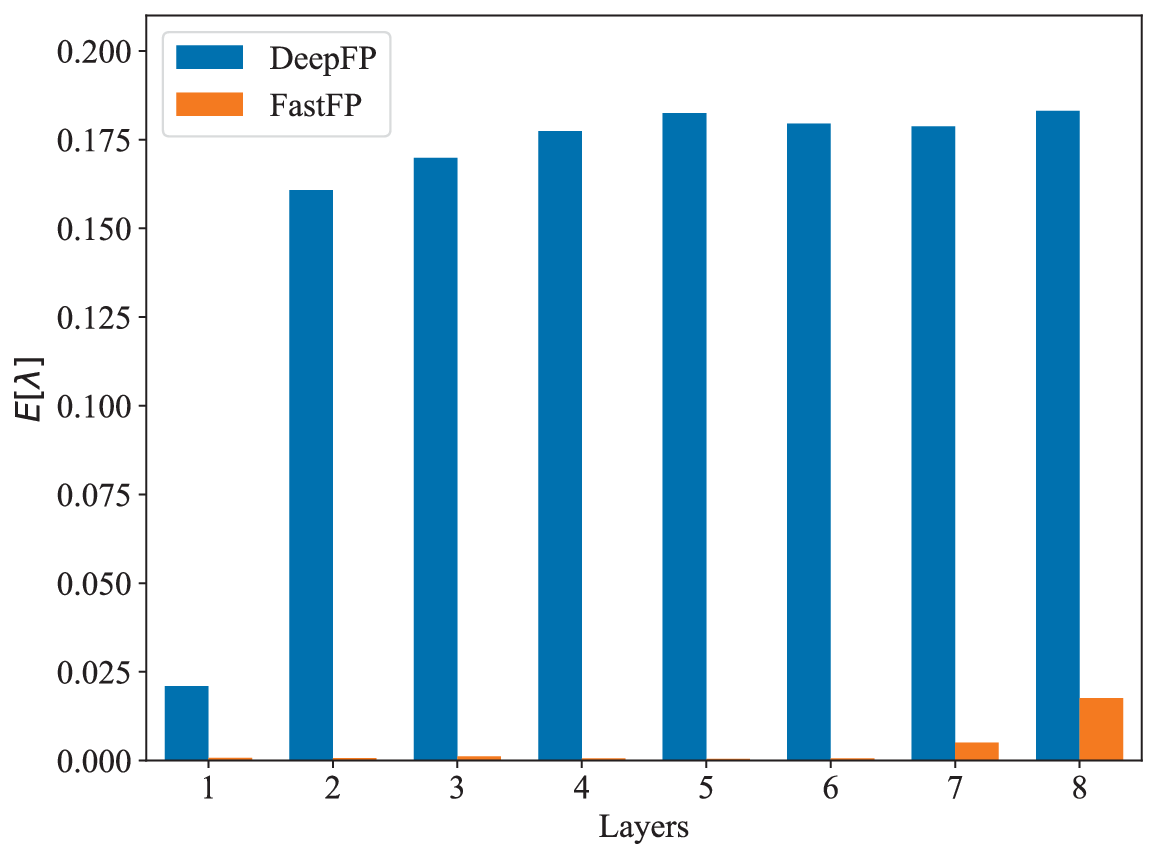}
        \caption{The mean of $\lambda$ selected by the FastFP algorithm and the DeepFP network. The FastFP algorithm computes $\lambda$  based on \eqref{equ:lambda_compute}, while $\lambda$ in the the DeepFP network is determined by the DNNs.}\label{fig:exp_lambda_compare}
\end{figure}

\begin{figure}[!t]
        \centering
        \includegraphics[width=0.4\textwidth]{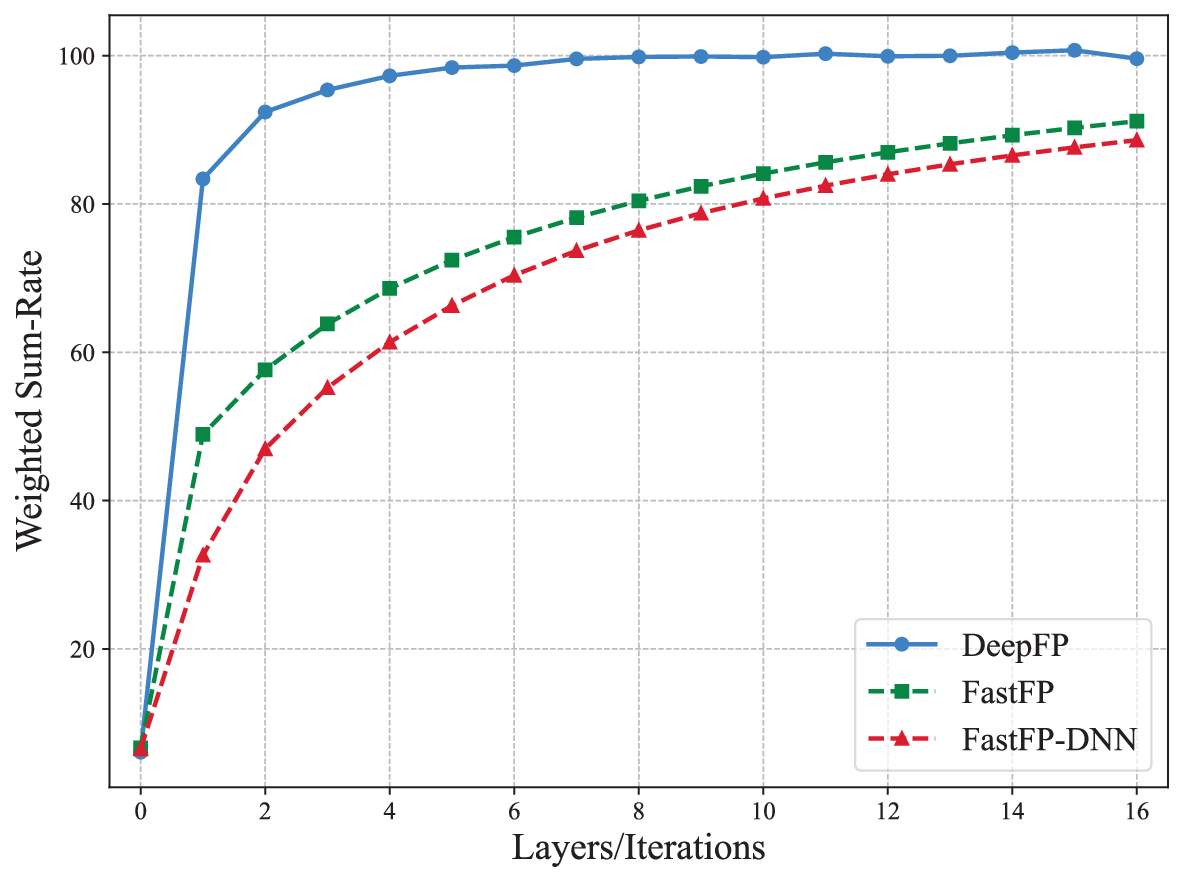}
        \caption{Weighted sum-rate performance of the DeepFP network, the FastFP algorithm, and the FastFP-DNN algorithm.}\label{fig:exp_WSR_fastfp_dnn}
\end{figure}

\begin{table}[!tbp]
\renewcommand\arraystretch{1.2}
        \centering
        \caption{Weighted Sum-Rate Performance of the DeepFP network with $N_r=4,d=2,K=6$ for different $N_t$. The DeepFP network is trained with $N_t=64$}\label{tab:gene_Nt}
        \begin{tabular}{|c|c|c|c|c|c|c|}
            \hline
                $N_t$ & Weighted Sum-Rate (bit/sec.) & Iterations by FastFP \\
            \hline
            $16$ & 55.224 (96.3\%)& 27\\
            $24$ & 67.962 (96.7\%)& 39\\
            $32$ & 77.502 (96.8\%)& 44\\
            $40$ & 82.584 (96.5\%)& 42\\
            $48$ & 86.880 (96.9\%)& 48\\
            $56$ & 93.546 (96.8\%)& 49\\
            $64$ & 99.474 (97.4\%)& 56\\
            \hline
        \end{tabular}
\end{table}

\subsubsection{Multicell Performance}

We further validate the WSR performance of the DeepFP network in a 7-cell wrapped-around network. The settings are $N_t = 64$, $N_r = 4$, $K = 6$, and $d = 2$. The FastFP algorithm and  the GCN-WMMSE algorithm~\cite{Schynol2023DUMIMO} serves as two baselines. We reimplemented the GCN-WMMSE algorithm in PyTorch and adjusted the size of its graph neural network to match that of the DNN in DeepFP for a fair complexity comparison. Table~\ref{tab:multicell} presents the WSR performance and CPU inference runtime. The results show that the proposed DeepFP network achieves 97.4\% of the WSR attained by FastFP while requiring only 11.7\% of its runtime. After $56$ iterations, FastFP achieves the same performance as the DeepFP network. Since GCN-WMMSE is designed to mimic the performance of the traditional WMMSE algorithm, it inherently inherits the associated weaknesses, such as slow convergence. Compared to GCN-WMMSE, DeepFP not only achieves higher WSR performance but also incurs lower inference latency.Fig.\ref{fig:exp_multicell_hist} and Fig.\ref{fig:exp_multicell_cdf} show the distribution and CDF of the WSR. The results indicate that the DeepFP network closely approximates the distribution of FastFP in the multicell MIMO system. Fig.~\ref{fig:exp_lambda_compare} shows the mean of $\lambda$ provided by the DNNs, as well as the mean of $\lambda$ calculated using  \eqref{equ:lambda_compute} under the same inputs. The results match our expectations: the DeepFP network produces smaller $\lambda$ values.

To enhance the fairness of the experiments, we designed a new baseline algorithm, named FastFP-DNN. Specifically, we used a DNN to directly predict the maximum eigenvalue in FastFP, and designed the DNN size to ensure that it has identical computational complexity to DeepFP, so that the complexity is well normalized. The results shown in Fig.~\ref{fig:exp_WSR_fastfp_dnn} demonstrate that the performance of the comparative algorithm is inferior to FastFP, while DeepFP achieves significantly superior performance compared to FastFP.

\begin{figure*}[t]
        \begin{minipage}{1\linewidth}
            \centering
            \subfloat[$K=6$]{\includegraphics[width=.33\linewidth]{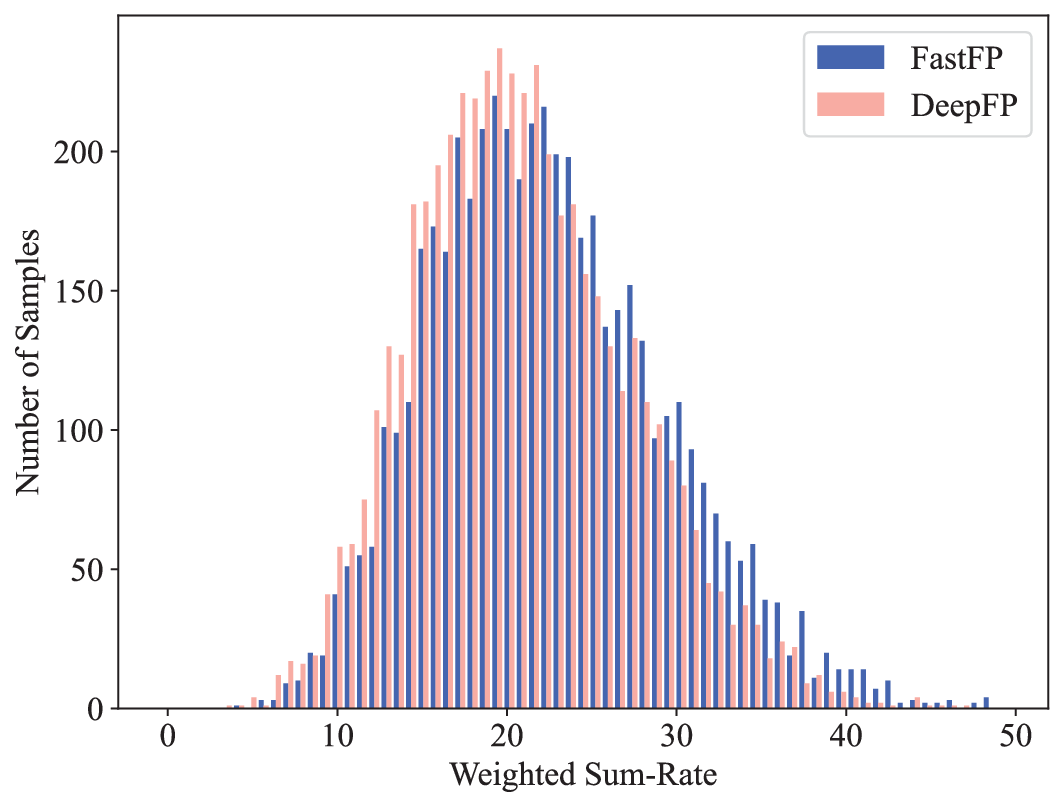}}
            \subfloat[$K=9$]{\includegraphics[width=.33\linewidth]{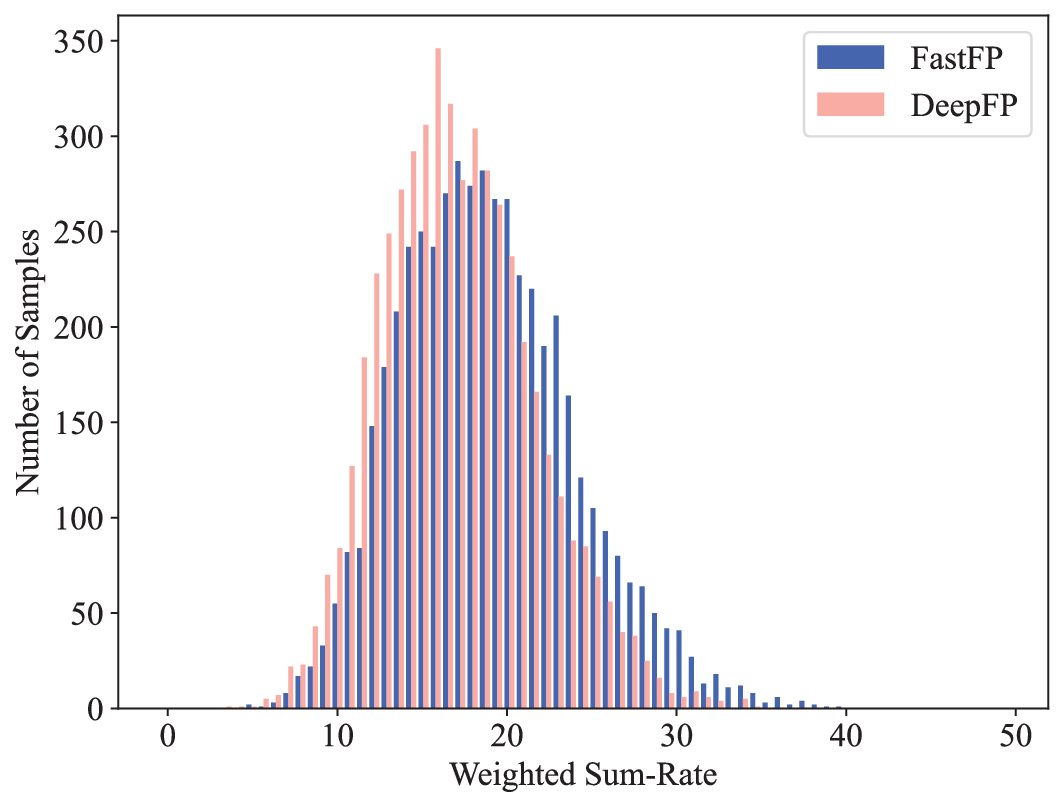}}
            \subfloat[$K=15$]{\includegraphics[width=.33\linewidth]{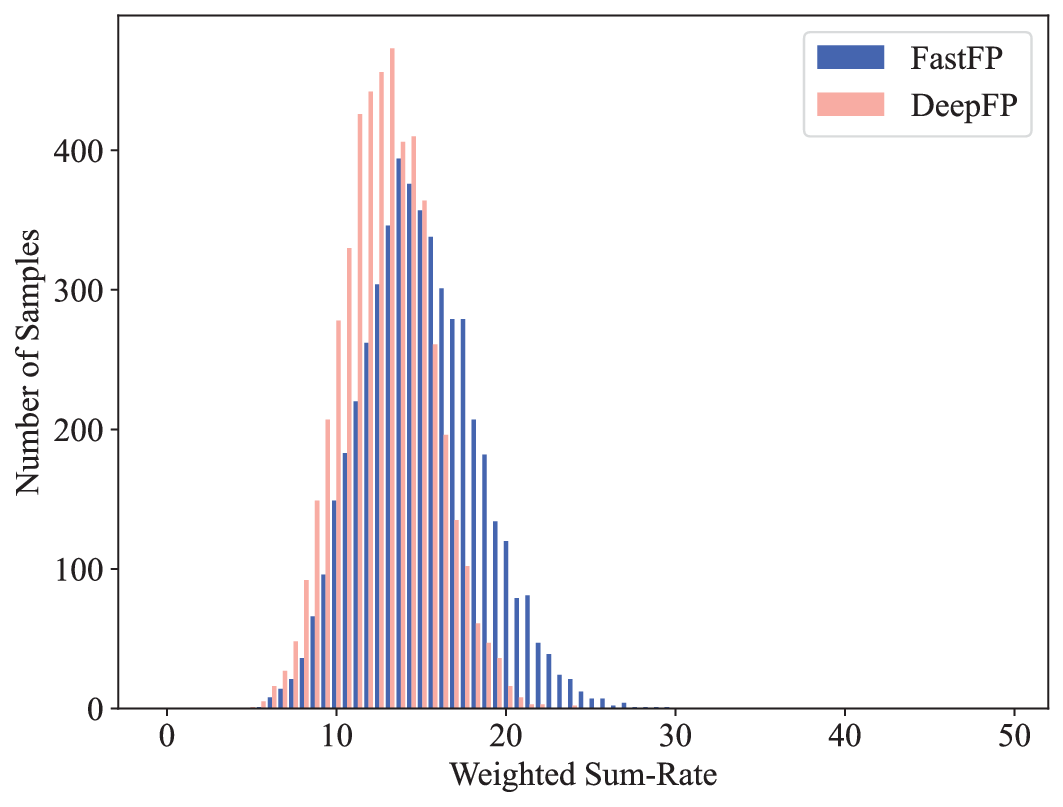}}
        \end{minipage}
        \caption{Distributions of the DeepFP network and the FastFP algorithm in 7-cell MIMO with $N_t=64,N_r=4,d=4$ for different $K$.}
        \label{fig:exp_multicell2_hist}
\end{figure*}

\begin{figure}[t]
        \centering
        \includegraphics[width=0.4\textwidth]{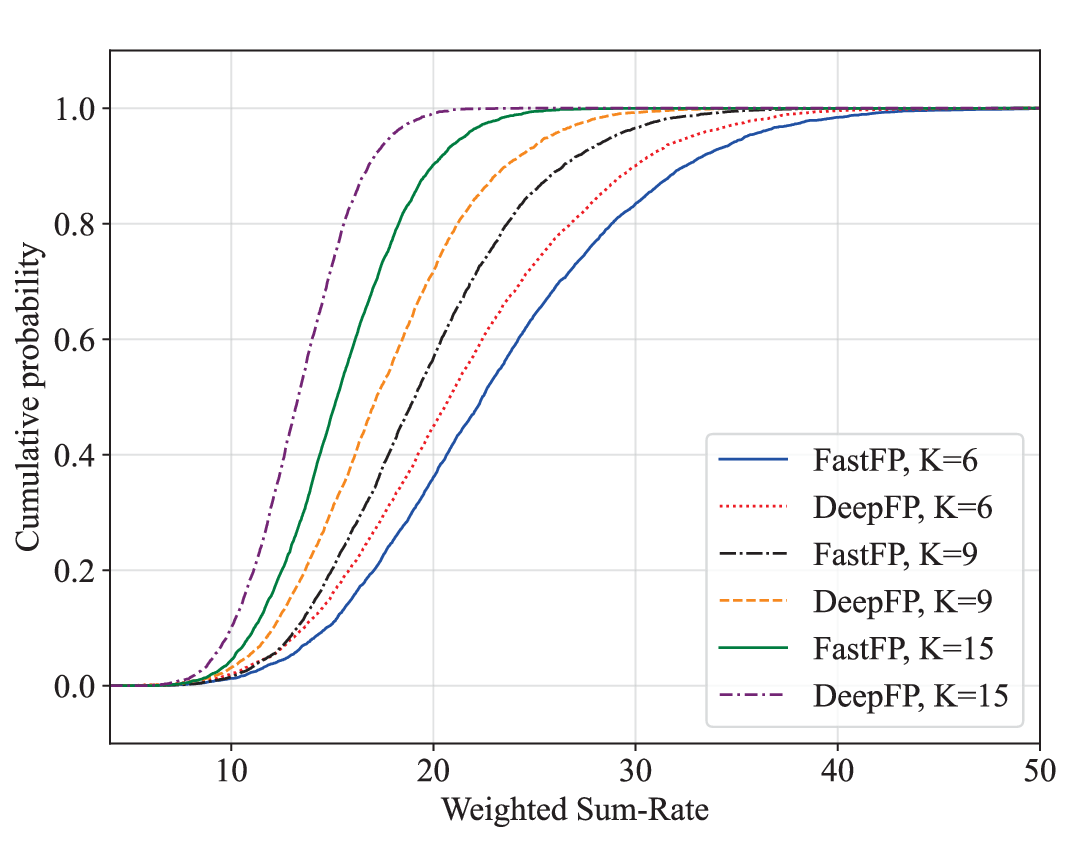}
        \caption{The CDF that describes the rates achieved by the DeepFP network and the FastFP algorithm in 7-cell MIMO with $N_t=64,N_r=4,d=4$ for different $K$.}
        \label{fig:exp_multicell2_cdf}
\end{figure}

Next, we consider scenarios with more users and higher data streams per user. We set $N_t = 64$, $N_r = 4$, $d = 4$, and $K = 6, 9, 15$. Table~\ref{tab:Kresults} presents the average WSR performance and CPU time. We define "Iterations by FastFP" as the average number of iterations FastFP requires to achieve the same performance as the DeepFP network. The results show that as the number of users increases, the WSR performance improves. However, the gap between the DeepFP network and FastFP also widens. Comparing Table~\ref{tab:Kresults} with Table~\ref{tab:multicell}, when $N_t = 64$, $N_r = 4$, and $K = 6$, the DeepFP network demonstrates better acceleration performance at $d = 2$. Fig.\ref{fig:exp_multicell2_hist} and Fig.\ref{fig:exp_multicell2_cdf} show the distribution and CDF of the WSR for different values of $K$, respectively. The results indicate that although the WSR performance of the DeepFP network decreases in percentage terms with an increasing number of users, it still provides a good approximation of the distribution of the FastFP algorithm.

\subsection{Generalizability Validation}

In the previous subsection, we evaluated the WSR and acceleration performance of the DeepFP network in MIMO systems of varying sizes. In practice, the test data often differs significantly from the training data, which arises from two aspects: 1) The test data may differ in size from the training data. For instance, in multicell MIMO, the number of users may change due to mobility. Additionally, we expect the trained network to be applicable to data from different cells, leading to variations in the number of transmit antennas. 2) Changes in the data distribution. Even if the test data has the same size as the training data, its distribution may differ. Therefore, in this subsection, we assess the generalization performance of the DeepFP network.

First, we use the network trained with $N_t = 64$, $N_r = 4$, $K = 6$, and $d = 4$ to test its performance under different values of $d$. The results are shown in Table~\ref{tab:gene_d}. These results indicate that the trained DeepFP network still performs well in terms of WSR for different values of $d$. As $d$ decreases, the number of iterations required by FastFP to achieve the same performance increases. Comparing the results for $d = 2$ with those in Table~\ref{tab:multicell}, the network's WSR performance decreases by $2.4\%$ when the number of data streams per user increases.

Next, we test the performance of the DeepFP network, trained with $N_t = 64$, $N_r = 4$, $K = 6$, and $d = 2$, under different values of $N_t$. The results are shown in Table~\ref{tab:gene_Nt}. These results indicate that as $N_t$ changes, the WSR performance of the DeepFP network remains stable at around $97\%$.

\begin{figure}[t]
        \centering
        \includegraphics[width=0.4\textwidth]{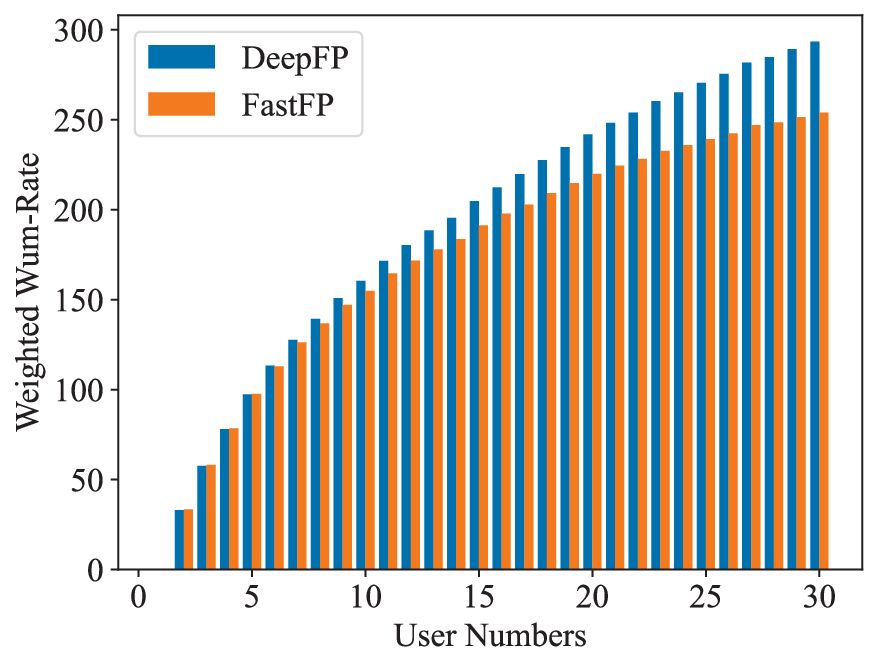}
        \caption{The WSR performance of the DeepFP network and the FastFP algorithm in 7-cell MIMO with $N_t=64,N_r=4,d=2$ for different $K$. The DeepFP network is trained with $N_t=64,N_r=4,d=2,K=6$.}
        \label{fig:gene_usernumber}
\end{figure}

\begin{table*}[t]
\renewcommand\arraystretch{1.2}
        \centering
        \caption{Weighted Sum-Rate Performance of the DeepFP Network for Different Cell Distances $D$ and Standard Deviation of $\xi$: The network is trained for $D = 0.8$ and a standard deviation of $\xi = 8$.}
        \label{tab:gene_distribution}
        \begin{tabular}{|c|ccc|ccc|}
        \hline
        \multirow{2}{*}{Cell Distance (km)} & 
        \multicolumn{3}{c|}{Weighted Sum-Rate (bit/sec.)} & 
        \multicolumn{3}{c|}{Iterations by FastFP} \\ 
        \cline{2-7}
         &  $\xi$ 4 dB & $\xi$ 8 dB & $\xi$ 12 dB  & $\xi$ 4dB & $\xi$ 8 dB & $\xi$ 12 dB \\ 
        \hline
        0.4 & 261.168 (93.2\%) & 261.012 (101.5\%) & 259.386 (113.2\%) & 58 & 132 & 233 \\
        0.6 & 167.178 (94.8\%) & 186.222 (99.1\%) & 205.962 (109.5\%) & 45 & 91 & 197 \\
        0.8 & 108.090 (96.1\%) & 136.746 (99.2\%) & 162.756 (107.5\%) & 35 & 89 & 152 \\
        1.0 & 73.386 (96.5\%) & 100.506 (99.1\%) & 127.536 (105.2\%) & 25 & 82 & 217 \\
        1.2 & 52.062 (97.0\%) & 76.446 (98.8\%) & 105.948 (103.5\%) & 21 & 71 & 204 \\
        \hline
        \end{tabular}
\end{table*}

Next, we continue using the DeepFP network trained with $N_t = 64$, $N_r = 4$, $K = 6$, and $d = 2$ to test its WSR performance on datasets with varying numbers of users. The results are shown in Fig.~\ref{fig:gene_usernumber}. These results indicate that when $K < 6$, the DeepFP network outperforms the FastFP algorithm after 100 iterations. As $K$ increases, the gap between the DeepFP network and the FastFP algorithm widens.

In the previous generalization test, we evaluated the trained network's generalization ability on test data of different sizes. Next, we test the network's generalization performance on data with different distributions. During the generation of the training data, the distance between base stations is set to $D = 0.8$ km, and the standard deviation of the path loss parameter $\xi$ is $8$ dB. We generate test data with varying values of $D$ and standard deviation, and then evaluate the network's WSR performance. The results are presented in Table~\ref{tab:gene_distribution}. These results show that the DeepFP network demonstrates good WSR performance under different distributions. To further assess generalization under distinct channel statistics, we evaluate the network in a Rayleigh fading scenario, trained under shadowing conditions. As shown in Table~\ref{tab:rayleigh_generalization}, the DNN trained for random shadowing can still yield comparable performance in the Rayleigh fading model.

\begin{table}[t]
\renewcommand\arraystretch{1.2}
\centering
\caption{Sum-Rate Performance of DeepFP and FastFP in a Rayleigh Fading Channel: The network is trained under a shadowing model and tested on Rayleigh fading without shadowing.}
\label{tab:rayleigh_generalization}
\begin{tabular}{|c|c|c|}
\hline
\multirow{2}{*}{Transmit Power (dB)} & 
\multicolumn{2}{c|}{Weighted Sum-Rate (bit/sec.)} \\ 
\cline{2-3}
 & FastFP & DeepFP \\ 
\hline
0 & 56.7086 & 68.4735 \\
10 & 62.5130 & 70.0218 \\
20 & 66.9725 & 70.4233 \\
30 & 71.3865 & 71.4206 \\
40 & 76.2021 & 73.0454 \\
\hline
\end{tabular}
\end{table}

\section{Conclusion}\label{sec:conclusion}

This work aims at a novel deep unfolding paradigm for optimizing the multicell MIMO beamformers in cellular networks. The proposed DeepFP method can be distinguished from the existing deep unfolding methods \cite{Hu2021IAIDNN, Minghe2023Learning, Pellaco2022UFWMMSE,Nguyen2023DUHybridbeamforming,Hu2021DURLMIMO,Xu2023JointSchebeamforming} for MIMO beamforming in two respects. First, while the previous work \cite{Xiaotong2023RWMMSE} can only reduce the complexity of large matrix inversion, DeepFP eliminates the large matrix inversion completely. Second, while  the previous work can linearize the Lagrange multiplier optimization only for a single cell, DeepFP extends the linearization for a generic multicell network. DeepFP acquires the above two benefits by linking the traditional WMMSE algorithm \cite{Shi2011WMMSE,Christensen2008WMMSE} with the FP tools \cite{kaiming2018FPI,kaiming2018FP2} and further incorporating an inhomogeneous bound \cite{Kaiming2024NFP} into the DNN design for deep unfolding. Extensive numerical examples show that DeepFP reduces the complexity of the conventional model-driven iterative algorithms (such as WMMSE) and can even outperform them in maximizing the WSR for multicell MIMO networks.

% \section*{Acknowledgments}
% This should be a simple paragraph before the References to thank those individuals and institutions who have supported your work on this article.

%{\appendices
%\section*{Proof of the First Zonklar Equation}
%Appendix one text goes here.
% You can choose not to have a title for an appendix if you want by leaving the argument blank
%\section*{Proof of the Second Zonklar Equation}
%Appendix two text goes here.}

 % argument is your BibTeX string definitions and bibliography database(s)

\bibliographystyle{IEEEtran}
\bibliography{IEEEabrv,refer}

@STRING{IEEE_J_SP         = "{IEEE} Trans. Signal Process."}

@STRING{IEEE_J_JSAC       = "{IEEE} J. Sel. Areas Commun."}

@STRING{IEEE_J_WCOM       = "{IEEE} Trans. Wireless Commun."}

@STRING{IEEE_ICASSP        = "Proc. {IEEE} Int. Conf. Acoust., Speech, Signal Process. (ICASSP)"}

@STRING{IEEE_WCNC        = "Proc. {IEEE} Wireless Commun. Netw. Conf. (WCNC)"}

@STRING{ICML          = "Int. Conf. Mach. Learn. (ICML)"}

@book{Bertsekas1999,
  author    = {D. P. Bertsekas},
  title     = {{Nonlinear Programming}},
  publisher = {MIT Press},
  address   = {Cambridge, MA, USA},
  year      = {1999}
}

@inproceedings{Zhu2025DeepFP,
  author    = {Jianhang Zhu and Tsung-Hui Chang and Liyao Xiang and Kaiming Shen},
  title     = {{DeepFP}: Deep-Unfolded Fractional Programming for Massive {MIMO} Beamforming},
  booktitle = {Proc. IEEE Int. Workshop Signal Process. Adv. Wireless Commun. (SPAWC)},
  month     = {June},
  year      = {2025}
}

@book{goldsmith2005wireless,
  title     = {{Wireless Communications}},
  author    = {Goldsmith, A.},
  year      = {2005},
  publisher = {Cambridge University Press},
  address   = {Cambridge, U.K.}
}

@ARTICLE{hojatian2021unsupervised,
  author={Hojatian, Hamed and Nadal, Jérémy and Frigon, Jean-François and Leduc-Primeau, François},
  journal={IEEE Trans. Wireless Commun.}, 
  title={{Unsupervised deep learning for massive MIMO hybrid beamforming}}, 
  year={2021},
  month = May,
  volume={20},
  number={11},
  pages={7086-7099}
}

@INPROCEEDINGS{Gregor2010LISTA,
  author={K. Gregor and Y. LeCun},
  booktitle=ICML,
  title={Learning fast approximations of sparse coding},
  year={2010},
  volume={},
  number={},
month = {June},
  pages={399-406}
}

@article{Christensen2008WMMSE,
  author   = {Christensen, Søren Skovgaard and Agarwal, Rajiv and De Carvalho, Elisabeth and Cioffi, John M.},
  journal  = IEEE_J_WCOM,
  title    = {{Weighted sum-rate maximization using weighted MMSE for MIMO-BC beamforming design}},
  year     = {2008},
  volume   = {7},
  number   = {12},
  month    = {Dec.},
  pages    = {4792-4799}
}

@inproceedings{Stimming2019DUCom,
  author    = {Balatsoukas-Stimming, Alexios and Studer, Christoph},
  booktitle = {2019 IEEE Workshop Signal Process. Syst. (SiPS)},
  title     = {{Deep unfolding for communications systems: A survey and some new directions}},
  year      = {2019},
  volume    = {},
  number    = {},
month = {Mar.},
  pages     = {266-271}
}

@inproceedings{Gao2011ZF,
  author    = {Gao, Xiang and Edfors, Ove and Rusek, Fredrik and Tufvesson, Fredrik},
  booktitle = {2011 IEEE Veh. Technol. Conf. (VTC Fall)},
  title     = {{Linear pre-coding performance in measured very-large MIMO channels}},
  year      = {2011},
  volume    = {},
  number    = {},
    month = {Dec.},
  pages     = {1-5}
}

@article{Haoran2018L2O,
  author  = {Sun, Haoran and Chen, Xiangyi and Shi, Qingjiang and Hong, Mingyi and Fu, Xiao and Sidiropoulos, Nicholas D.},
  journal = IEEE_J_SP,
  title   = {{Learning to optimize: Training deep neural networks for interference management}},
  year    = {2018},
  volume  = {66},
  number  = {20},
month = {Oct.},
  pages   = {5438-5453}
}

@article{hengtao2018DLMIMO,
  author  = {He, Hengtao and Wen, Chao-Kai and Jin, Shi and Li, Geoffrey Ye},
  journal = {IEEE Wireless Commun. Lett.},
  title   = {{Deep learning-based channel estimation for beamspace mmWave massive MIMO systems}},
  year    = {2018},
  volume  = {7},
  number  = {5},
month = {Oct},
  pages   = {852-855}
}

@misc{hershey2014deepunfolding,
  title         = {{Deep unfolding: Model-based inspiration of novel deep architectures}},
  author        = {John R. Hershey and Jonathan Le Roux and Felix Weninger},
  year          = {2014},
  eprint        = {1409.2574},
  archiveprefix = {arXiv},
  primaryclass  = {cs.LG},
  url           = {https://arxiv.org/abs/1409.2574}
}

@article{wenchao2020DLMISO,
  author  = {Xia, Wenchao and Zheng, Gan and Zhu, Yongxu and Zhang, Jun and Wang, Jiangzhou and Petropulu, Athina P.},
  journal = {IEEE Trans. Commun.},
  title   = {{A deep learning framework for optimization of MISO downlink beamforming}},
  year    = {2020},
  volume  = {68},
  number  = {3},
month = {Mar.},
  pages   = {1866-1880}
}

@article{wei2018DL,
  author  = {Cui, Wei and Shen, Kaiming and Yu, Wei},
  journal = {IEEE J. Sel. Areas Commun.},
  title   = {{Spatial deep learning for wireless scheduling}},
  year    = {2019},
  volume  = {37},
  number  = {6},
month = {June},
  pages   = {1248-1261}
}

@article{Hu2021IAIDNN,
  author  = {Hu, Qiyu and Cai, Yunlong and Shi, Qingjiang and Xu, Kaidi and Yu, Guanding and Ding, Zhi},
  journal = {IEEE Trans. Wireless Commun.},
  title   = {{Iterative algorithm induced deep-unfolding neural networks: Precoding design for multiuser MIMO systems}},
  year    = {2021},
  volume  = {20},
  number  = {2},
month = {Feb.},
  pages   = {1394-1410}
}

@inproceedings{Joshi2011WSRBB,
  author    = {Joshi, S. and Weeraddana, P. C. and Codreanu, M. and Latva-aho, M.},
  booktitle = {2011 Conf. Rec. Forty Fifth Asilomar Conf. Signals, Syst. Comput. (ASILOMAR)},
  title     = {{Weighted sum-rate maximization for MISO downlink cellular networks via branch and bound}},
  year      = {2011},
  volume    = {},
  number    = {},
month  = {Apr.},
  pages     = {1569-1573}
}

@article{kaiming2018FPI,
  author  = {Shen, Kaiming and Yu, Wei},
  journal = {IEEE Trans. Signal Process.},
  title   = {{Fractional programming for communication systems—{Part} I: Power control and beamforming}},
  year    = {2018},
  volume  = {66},
  number  = {10},
  month = {May},
  pages   = {2616-2630}
}

@article{kaiming2018FP2,
  author  = {Shen, Kaiming and Yu, Wei},
  journal = {IEEE Trans. Signal Process.},
  title   = {{Fractional programming for communication systems—{Part} II: Uplink scheduling via matching}},
  year    = {2018},
  volume  = {66},
  number  = {10},
month = {May},
  pages   = {2631-2644}
}

@article{Xu2023JointSchebeamforming,
  author  = {Xu, Chunmei and Jia, Yuanqi and He, Shiwen and Huang, Yongming and Niyato, Dusit},
  journal = {IEEE Trans. Commun.},
  title   = {{Joint user scheduling, base station clustering, and beamforming design based on deep unfolding technique}},
  year    = {2023},
  volume  = {71},
  number  = {10},
month = {Oct.},
  pages   = {5831-5845}
}

@article{Hu2021DURLMIMO,
  author  = {Hu, Qiyu and Liu, Yanzhen and Cai, Yunlong and Yu, Guanding and Ding, Zhi},
  journal = {IEEE J. Sel. Areas Commun.},
  title   = {{Joint deep reinforcement learning and unfolding: Beam selection and precoding for mmWave multiuser MIMO with lens arrays}},
  year    = {2021},
  volume  = {39},
  number  = {8},
 month = {Aug.},
  pages   = {2289-2304}
}

@article{Nguyen2023DUHybridbeamforming,
  author  = {Nguyen, Nhan Thanh and Ma, Mengyuan and Lavi, Ortal and Shlezinger, Nir and Eldar, Yonina C. and Swindlehurst, A. Lee and Juntti, Markku},
  journal = {IEEE Trans. Signal Process.},
  title   = {{Deep unfolding hybrid beamforming designs for THz massive MIMO systems}},
  year    = {2023},
  volume  = {71},
  number  = {},
month={Oct.},
  pages   = {3788-3804}
}

@misc{ReLU,
  title         = {{Deep learning using rectified linear units (ReLU)}},
  author        = {Abien Fred Agarap},
  year          = {2019},
  eprint        = {1803.08375},
  archiveprefix = {arXiv},
  primaryclass  = {cs.NE},
  url           = {https://arxiv.org/abs/1803.08375}
}

@article{Xiaotong2023RWMMSE,
  author  = {Zhao, Xiaotong and Lu, Siyuan and Shi, Qingjiang and Luo, Zhi-Quan},
  journal = {IEEE Trans. Signal Process.},
  title   = {{Rethinking WMMSE: Can its complexity scale linearly with the number of BS antennas?}},
  year    = {2023},
  volume  = {71},
  number  = {},
month = {Feb.},
  pages   = {433-446}
}

@article{Kaiming2024NFP,
  author     = {Shen, Kaiming and Zhao, Ziping and Chen, Yannan and Zhang, Zepeng and V. Cheng, H.},
  title      = {{Accelerating quadratic transform and WMMSE}},
  year       = {2024},
  issue_date = {Nov. 2024},
  volume     = {42},
  number     = {11},
  issn       = {0733-8716},
  journal    = {IEEE J. Sel. Areas Commun.},
  month      = {July},
  pages      = {3110–3124}
}

@article{Kammoun2014MRT,
  author  = {Kammoun, Abla and Müller, Axel and Björnson, Emil and Debbah, Mérouane},
  journal = {IEEE J. Sel. Top. Signal Process.},
  title   = {{Linear precoding based on polynomial expansion: Large-scale multi-cell MIMO systems}},
  year    = {2014},
  volume  = {8},
  number  = {5},
month = {Oct.},
  pages   = {861-875}
}

@inproceedings{Rui2024Manifold,
  author    = {Sun, Rui and Wang, Chen and Lu, An-An and Fu, Xiao and Liu, Xiaofeng and Zhang, Yuxuan and Gao, Xiqi and Xia, Xiang-Gen},
  booktitle = IEEE_WCNC,
  title     = {{Matrix manifold precoder design for massive MIMO downlink}},
  year      = {2024},
  volume    = {},
  number    = {},
month = {July},
  pages     = {1-6}
}

@INPROCEEDINGS{Zhang2023Enhangcing,
  author={Zhang, Zepeng and Zhao, Ziping and Shen, Kaiming},
  booktitle=IEEE_ICASSP, 
  title={Enhancing the efficiency of {WMMSE} and {FP} for beamforming by Minorization-maximization}, 
  year={2023},
  volume={},
  number={},
month = {May}
}

@article{Liu2012WSRBB,
  author  = {Liu, Liang and Zhang, Rui and Chua, Kee-Chaing},
  journal = {IEEE Trans. Wireless Commun.},
  title   = {{Achieving global optimality for weighted sum-rate maximization in the $K$-user Gaussian interference channel with multiple antennas}},
  year    = {2012},
  volume  = {11},
  number  = {5},
month = { May},
  pages   = {1933-1945}
}

@article{luo2008Dynamic,
  author  = {Luo, Zhi-Quan and Zhang, Shuzhong},
  journal = {IEEE J. Sel. Top. Signal Process.},
  title   = {{Dynamic spectrum management: Complexity and duality}},
  year    = {2008},
  volume  = {2},
  number  = {1},
month  = {Feb.},
  pages   = {57-73}
}

@article{Minghe2023Learning,
  author  = {Zhu, Minghe and Chang, Tsung-Hui and Hong, Mingyi},
  journal = {IEEE Trans. Wireless Commun.},
  title   = {{Learning to beamform in heterogeneous massive MIMO networks}},
  year    = {2023},
  volume  = {22},
  number  = {7},
month = {July},
  pages   = {4901-4915}
}

@article{Nguyen2019RZF,
  author  = {Nguyen, Long D. and Tuan, Hoang Duong and Duong, Trung Q. and Poor, H. Vincent},
  journal = {IEEE Trans. Signal Process.},
  title   = {{Multi-user regularized zero-forcing beamforming}},
  year    = {2019},
  volume  = {67},
  number  = {11},
month = {June},
  pages   = {2839-2853}
}

@article{Pellaco2022UFWMMSE,
  author  = {Pellaco, Lissy and Bengtsson, Mats and Jaldén, Joakim},
  journal = {IEEE Open J. Commun. Soc.},
  title   = {{Matrix-inverse-free deep unfolding of the weighted MMSE beamforming algorithm}},
  year    = {2022},
  volume  = {3},
  number  = {},
    month = {Dec.},
  pages   = {65-81}
}

@article{Shi2011WMMSE,
  author  = {Shi, Qingjiang and Razaviyayn, Meisam and Luo, Zhi-Quan and He, Chen},
  journal = {IEEE Trans. Signal Process.},
  title   = {An iteratively weighted {MMSE} approach to distributed sum-utility maximization for a {MIMO} interfering broadcast channel},
  year    = {2011},
  volume  = {59},
  month   = {Sept.},
  number  = {9},
  pages   = {4331-4340}
}

@article{Sunying2017MM,
  author   = {Sun, Ying and Babu, Prabhu and Palomar, Daniel P.},
  journal  = {IEEE Trans. Signal Process.},
  title    = {{Majorization-minimization algorithms in signal processing, communications, and machine learning}},
  year     = {2017},
  volume   = {65},
  number   = {3},
  pages    = {794-816},
month = {Feb.},
  doi      = {10.1109/TSP.2016.2601299}
}

@article{Schynol2023DUMIMO,
  author  = {Schynol, Lukas and Pesavento, Marius},
  journal = IEEE_J_JSAC,
  title   = {{Coordinated sum-rate maximization in multicell MU-MIMO with deep unrolling}},
  year    = {2023},
  volume  = {41},
  number  = {4},
month = {Apr.},
  pages   = {1120-1134}
}

@article{Chowdhury2024GNNMIMO,
  author  = {Chowdhury, Arindam and Verma, Gunjan and Swami, Ananthram and Segarra, Santiago},
  journal = IEEE_J_WCOM,
  title   = {{Deep graph unfolding for beamforming in MU-MIMO interference networks}},
  year    = {2024},
  volume  = {23},
  number  = {5},
month  ={May},
  pages   = {4889-4903}
}

\vspace{11pt}

\newpage

% \bf{If you include a photo:}\vspace{-33pt}
\begin{IEEEbiographynophoto}
{Jianhang Zhu} (Student Member, IEEE) 
received the B.E. degree in communication engineering from the School of Electronics and Information Technology, Sun Yat-sen University, Guangzhou, China, in 2021, and the M.E. degree in computer science from the School of Computer Science and Engineering, Sun Yat-sen University, in 2024. He is currently pursuing the Ph.D. degree with the School of Science and Engineering, The Chinese University of Hong Kong (Shenzhen), China. His research interests include semantic communication, deep unfolding, non-convex optimization, machine learning, the Age of Information, edge computing, and the Internet of Things.
\end{IEEEbiographynophoto}

\begin{IEEEbiographynophoto}
{Tsung-Hui Chang} (Fellow, IEEE) received the B.S. degree in electrical engineering and the Ph.D. degree in communications engineering from the National Tsing Hua University (NTHU), Hsinchu, Taiwan, in 2003 and 2008, respectively. Currently, he is a Professor and the Associate Dean (Education) of the School of Artificial Intelligence, The Chinese University of Hong Kong, Shenzhen (CUHK-SZ), China, and the Associate Director of Guangdong Provincial Key Laboratory of Big Data Computing. Before joining CUHK-SZ, he was with the National Taiwan University of Science and Technology (NTUST), the University of California, Davis, as a Postdoctoral Researcher and a Faculty Member, respectively. His research interests include signal processing and optimization problems in data communications and machine learning. He has been an Elected Member of the IEEE Signal Processing Society (SPS) Signal Processing for Communications and Networking Technical Committee (SPCOM TC) since 2020. He received the Young Scholar Research Award of NTUST in 2014, the IEEE ComSoc Asian-Pacific Outstanding Young Researcher Award in 2015, the IEEE SPS Best Paper Awards in 2018 and 2021, the Outstanding Faculty Research Award of SSE of CUHK-SZ in 2021, and the Outstanding Research Award of CUHK-SZ in 2024. He is the Founding Chair of the IEEE SPS Integrated Sensing and Communication Technical Working Group (ISAC TWG) and the elected Regional Director-atLarge of Board of Governors of IEEE SPS from 2022 to 2023. He has served on the editorial board for major SP journals, including an Associate Editor for IEEE TRANSACTIONS ON SIGNAL PROCESSING from 2014 to 2018, IEEE TRANSACTIONS ON SIGNAL AND INFORMATION PROCESSING OVER NETWORKS from 2015 to 2018, IEEE OPEN JOURNAL OF SIGNAL PROCESSING since 2020, and a Senior Area Editor for IEEE TRANSACTIONS ON SIGNAL PROCESSING from 2021 to 2025.
\end{IEEEbiographynophoto}

\begin{IEEEbiographynophoto}{Liyao Xiang} (Member, IEEE) received the B.Eng. degree in Electrical and Computer Engineering from Shanghai Jiao Tong University, Shanghai, China, in 2012, and the Ph.D. degree in Computer Engineering from the University of Toronto, Toronto, ON, Canada, in 2018. She is currently an associate professor at Shanghai Jiao Tong University. Her research interests include AI security and privacy, privacy analysis in data mining, and mobile computing.
\end{IEEEbiographynophoto}

\begin{IEEEbiographynophoto}{Kaiming Shen} (Senior Member, IEEE) received the B.Eng. degree in information security and the B.Sc. degree in mathematics from Shanghai Jiao Tong University, China in 2011, and then the M.A.Sc. degree in electrical and computer engineering from the University of Toronto, Canada in 2013. After working at a tech startup in Ottawa for one year, he returned to the University of Toronto and received the Ph.D. degree in electrical and computer engineering in early 2020. He has been with the School of Science and Engineering at The Chinese University of Hong Kong, Shenzhen, China as a tenure-track assistant professor since 2020. His research interests include optimization, wireless communications, and information theory. He currently serves as an Editor for IEEE Transactions on Wireless Communications. He is a member of the Signal Processing for Communications and Networking Technical Committee of the IEEE Signal Processing Society. He received the IEEE Signal Processing Society Young Author Best Paper Award in 2021, the University Teaching Achievement Award in 2023, the Frontiers of Science Award in 2024, and the Chinese Information Theory Society Young Researcher Award in 2025.
\end{IEEEbiographynophoto}

\vfill

\end{document}